\documentclass[prd,nofootinbib,notitlepage,10pt,aps]{revtex4-1}
\usepackage{hyperref,color,amsmath,amssymb,epsfig,graphicx}
\usepackage{epsfig,slashed}
\usepackage[utf8]{inputenc}
\usepackage{tensor}
\usepackage{tipa}      
\usepackage{textcomp}  
\usepackage{latexsym}  
\usepackage{amsfonts}
\usepackage{array}
\usepackage{bigstrut}
\usepackage{bm,bbold}

\newcommand{\media}[1]{\langle #1 \rangle}
\newcommand{\di}{{\rm d}}
\newcommand{\ii}{ \mathrm{i}}

\newcommand{\h}[1]{\widehat{#1}}
\def\spt{{\cal S}}
\def\wT{{\widehat T}}

\def\wspt{{\widehat{\cal S}}}

\def\wpsi{{\widehat{\psi}}}
\def\wbPsi{{\sf{\bar\Psi}}}
\def\wPsi{{\sf{\Psi}}}
\def\wrho{{\widehat{\rho}}}

\newcommand{\tr}{{\rm tr}}  
  
\newcommand{\e}{{\rm e}}

\newcommand{\x}{{\rm x}}

\newcommand{\de}{\partial}
\newcommand{\ped}[1]{_{\textup{#1}}}
\newcommand{\apic}[1]{^{\textup{#1}}}
\newcommand{\be}{\begin{equation}}
\newcommand{\ee}{\end{equation}}                          
\newcommand{\f}{\frac}   
\newcommand{\E}{{\rm e}}
\newcommand{\bea}{\begin{eqnarray}}
\newcommand{\eea}{\end{eqnarray}}                         
\DeclareMathOperator*{\SumInt}{%
\mathchoice%
  {\ooalign{$\displaystyle\sum$\cr\hidewidth$\displaystyle\int$\hidewidth\cr}}
  {\ooalign{\raisebox{.14\height}{\scalebox{.7}{$\textstyle\sum$}}\cr\hidewidth$\textstyle\int$\hidewidth\cr}}
  {\ooalign{\raisebox{.2\height}{\scalebox{.6}{$\scriptstyle\sum$}}\cr$\scriptstyle\int$\cr}}
  {\ooalign{\raisebox{.2\height}{\scalebox{.6}{$\scriptstyle\sum$}}\cr$\scriptstyle\int$\cr}}
}
\begin{document}

\title{General equilibrium second-order hydrodynamic coefficients for free quantum fields} 

\author{M. Buzzegoli}
\affiliation{Universit\'a di Firenze and INFN Sezione di Firenze,
Via G. Sansone 1, I-50019 Sesto Fiorentino (Firenze), Italy}
\author{E. Grossi}
\affiliation{Institut f\"ur Theoretische Physik, Universit\"at Heidelberg,
Philosophenweg 16, D-69120 Heidelberg, Germany}
\author{F. Becattini}
\affiliation{Universit\'a di Firenze and INFN Sezione di Firenze,
Via G. Sansone 1, I-50019 Sesto Fiorentino (Firenze), Italy\\
New York University Florence, Via Bolognese 120, I-50139, Firenze, Italy}

%

\begin{abstract}
We present a systematic calculation of the corrections of the stress-energy tensor 
and currents of the free boson and Dirac fields up to second order in thermal vorticity, 
which is relevant for relativistic hydrodynamics. These corrections are non-dissipative because they 
survive at general thermodynamic equilibrium with non vanishing mean values of the conserved 
generators of the Lorentz group, i.e. angular momenta and boosts. Their equilibrium nature  
makes it possible to express the relevant coefficients by means of correlators of the angular-momentum 
and boost operators with stress-energy tensor and current, thus making simpler to determine 
their so-called "Kubo formulae". We show that, at least for free fields, the corrections are
of quantum origin and we study several limiting cases and compare our results with previous 
calculations. We find that the axial current of the free Dirac field receives corrections
proportional to the vorticity independently of the anomalous term.
\end{abstract}
\maketitle

\section{Introduction}

Relativistic hydrodynamics is an effective dynamical theory of systems at local 
thermodynamic equilibrium. The condition of local thermodynamic equilibrium, makes it 
possible to describe the dynamics of interacting quantum fields with classical partial 
differential equations involving few thermodynamic fields (temperature, velocity, chemical 
potentials) provided that there is a separation between macroscopic and microscopic 
scales: the thermodynamic fields should vary significantly over distances which are 
much (or sufficiently) larger than microscopic scales of the quantum theory, such as mean 
free path, Compton wavelength etc. The branches of physics for which relativistic hydrodynamics 
is an effective tool include astrophysics, cosmology as well relativistic heavy ion collisions \cite{Baier:2006um,Huovinen:2006jp,Florkowski,Gale:2013da,Heinz:2013th,kodama}. In the past few years,
the derivation of hydrodynamic equations has drawn much attention: from kinetic theory \cite{York:2008rr,Muronga:2006zx,Tsumura:2006hn,Betz:2008me,Monnai:2010qp,Van:2011yn,denicol,
jaiswal}, from fluid/gravity correspondence \cite{baier,Bhattacharyya:2008jc,Natsuume:2007ty,
Hubeny:2011hd}, from the phenomenological extension of the non equilibrium thermodynamics 
\cite{Koide:2006ef,Fukuma:2011pr}, from the projection operator method \cite{Koide:2008nw,Minami:2012hs},
from non equilibrium statistical operator method \cite{Hayata:2015} and from imposing some symmetries 
on the system \cite{Banerjee:2012iz,kovtun}.
The success of relativistic hydrodynamics in heavy ion collisions lately arose fundamental 
questions about its domain of applicability \cite{Romatschke:2016hle}.

Relativistic hydrodynamics is based on the expansion of the mean stress-energy tensor as a function
of the thermodynamic fields and their gradients. The zero-order term is the well known ideal
form of the mean stress-energy tensor: 
\begin{equation}\label{tmunuideal}
T^{\mu\nu}=(\rho+p) u^\mu u^\nu-g^{\mu\nu}p,
\end{equation}
which is an exact expression at the global thermodynamic equilibrium with constant parameters 
$T,\mu$ and $u$. The first-order corrections to the (\ref{tmunuideal}) in the gradients of 
$u$, $T$ ad $\mu$ are dissipative, that is they imply the irreversible increase of entropy. The 
second-order corrections - including quadratic terms in first-order derivatives as well as 
second order derivatives - to the ideal form were classified in refs.~\cite{baier,Romatschke:2009kr,Moore:2010bu} 
and discussed in refs.~\cite{Bhattacharyya:2008jc,York:2008rr,moore,kovtun,Becattini:2015nva,Panerai}.

Among the second order terms in the gradients, there are quadratic terms in acceleration
and vorticity which do not contribute to the entropy increase, i.e. they are non-dissipative.
Such terms were dubbed in ref.~\cite{moore} as thermodynamic, and it was shown in ref.~\cite{Becattini:2015nva}
that they appear in the most general form of thermodynamic equilibrium in flat spacetime involving
non-vanishing acceleration and vorticity, such as the rotating fluid with constant angular 
velocity \cite{landau,vilenkin}. Furthermore, in ref.~\cite{Becattini:2015nva} was shown that, 
at least for free fields, these terms are of quantum origin, namely they vanish in the $\hbar \to 0$ 
limit, as it was already argued in ref.~\cite{Romatschke:2009kr}. Their quantum origin is also 
born out by the fact that they are missing in the stress-energy tensor expression from the 
classical expansion of the Boltzmann distribution function in the kinetic theory~\cite{jaiswal}.
The interest for such terms has been certainly reinforced by the recent observation of a
thermal vorticity of the order of some percent by the STAR experiment through the measurement 
of hyperon polarization \cite{star}.

The scalar coefficient multiplying the term $\omega^\mu\omega^\nu$, where $\omega$ is
the kinematic vorticity, was calculated in ref.~\cite{moore} with a suitable ``Kubo formula" 
involving stress-energy tensor three-point functions. 
In ref.~\cite{Becattini:2015nva} it was shown that, on the other hand, those coefficients 
can be expressed as correlators of conserved generators of the Poincar\'e group and the 
stress-energy tensor at the usual thermodynamic equilibrium with constant $T$ and $\mu$. 
This property involves a remarkable simplification of the "Kubo formulae", insofar as the 
constancy of generators allows to remove time integration, unlike in those of dissipative 
coefficients, like shear viscosity. 
The ultimate reason thereof is the relation of these terms to global equilibrium configurations.
In fact, in this paper we will use a different method, where the time integration survives
in imaginary time.

In ref.~\cite{Becattini:2015nva} their explicit expression was found for the real scalar 
free field. In this work, we extend that calculation to the free charged scalar field 
and the Dirac field including a finite chemical potential. We will be using an operator
formalism throughout instead of the functional approach~\cite{kovtun,Bhattacharyya:2012nq,
Banerjee:2012iz}. As it will be shown, the operator formalism is very convenient 
for a simple and systematic derivation of the "Kubo formulae" of the non-dissipative
coefficients. On the other hand, the functional approach allows to obtain relations 
between those coefficients \cite{kovtun,Bhattacharyya:2012nq,Banerjee:2012iz}, which 
are more difficult to extract in the operator formalims. 

\subsection*{Notation}

In this paper we use the natural units, with $\hbar=c=k_B=1$.\\ 
The Minkowskian metric tensor is ${\rm diag}(1,-1,-1,-1)$; for the Levi-Civita
symbol we use the convention $\epsilon^{0123}=1$.\\ 
We will use the relativistic notation with repeated indices assumed to be summed over, however 
contractions of indices will be sometimes denoted with dots, e.g. $ u \cdot T \cdot u \equiv u_\mu T^{\mu\nu} u_\nu$. Operators in Hilbert space will be denoted by a large upper hat, e.g. $\wT$ (with the 
exception of Dirac field operator that is denoted by $\wPsi$) while unit 
vectors with a small upper hat, e.g. $\hat v$. The stress-energy tensor is
assumed to be symmetric with an associated vanishing spin tensor.

\section{Global thermodynamic equilibrium density operator}

The general covariant form of the local thermodynamic equilibrium density operator was 
introduced in \cite{zubarev,weert,weldon} and lately discussed in detail in refs.~\cite{becacov,becalocal,Becattini:2015nva,Hayata:2015,Hongo:2016mqm}:
\begin{equation}\label{gener1}
  \wrho = \frac{1}{Z} \exp \left[ -\int_{\Sigma} \di \Sigma_\mu 
  \left( \h T^{\mu\nu}(x) \beta_\nu(x) - \zeta(x)\, \h j^\mu(x) \right) \right],
\end{equation}
where $\h T^{\mu\nu}$ and $\h j^\mu$ are the stress-energy tensor and a conserved current 
operators, $\beta^{\mu}$ is the four-temperature vector such that
$$
  T = \frac{1}{\sqrt{\beta^2}}
$$
is the proper comoving temperature and $\zeta$ is the ratio of comoving chemical 
potential and temperature $\zeta=\mu /T$; $\Sigma$ is a spacelike 3-D hypersurface.
This operator is obtained by maximizing the entropy with fixed mean energy, momentum 
and charge density \cite{becalocal}. If time-like, the four-temperature vector $\beta$ 
defines a local four-velocity of the fluid (the {\em $\beta$ frame} \cite{becalocal,Van:2013sma}):
\begin{equation}\label{fourv}
 u_\mu(x)=\frac{\beta_\mu(x)}{\sqrt{\beta^2}}, 
\end{equation}
and the magnitude of $\beta$ is the proper temperature measured by a thermometer moving
with four-velocity $u$.

If the four-temperature $\beta$ is a Killing vector \cite{becacov}:
\begin{equation}\label{killing}
\nabla_{\mu}\beta_{\nu}+\nabla_{\nu}\beta_{\mu}=0\qquad \nabla_{\mu}\zeta=0,
\end{equation}
and $\zeta$ is constant, the density operator~(\ref{gener1}) is a proper global thermodynamic
equilibrium density operator insofar as it becomes independent of the hypersurface $\Sigma$
with suitable conditions at a timelike boundary. The general solution of the 
eq.~(\ref{killing}) in Minkowsky space-time is known and can be expressed in terms of 
a constant  four-vector $b_{\mu}$ and a constant antisymmetric tensor $\varpi_{\mu\nu}$:
\begin{equation}\label{betakilling}
\beta_\mu(x)=b_\mu+\varpi_{\mu\nu}x^{\nu}\quad \quad\zeta=\mathrm{const.}
\end{equation}
The antisymmetric tensor $\varpi_{\mu\nu}$ is called \emph{thermal vorticity}; from the
above equation it turns out that it is the antisymmetric part of the gradient of $\beta$:
\begin{equation}\label{thvort}
\varpi_{\mu\nu}=-\frac{1}{2}(\partial_\mu \beta_\nu -\partial_\nu \beta_\mu ).
\end{equation}
Once the eq.~(\ref{betakilling}) is plugged into the eq.~(\ref{gener1}), the general expression of 
the equilibrium density operator in flat space-time is obtained \cite{Becattini:2015nva}. 
\begin{equation}\label{densop}
\h \rho =\frac{1}{Z} \exp\left[ -b_\mu \h P^\mu +\frac{1}{2}\varpi_{\mu\nu}\h J^{\mu\nu}+\zeta \h Q\right],
\end{equation}
where $\h P$ is the four momentum operator, $\h Q$ the conserved charge and $\h J^{\mu\nu}$ 
are the generators of the Lorentz transformations:
\begin{equation}
\h J^{\mu\nu}=\int_\Sigma \di \Sigma_{\lambda} 
\left( x^{\mu}\h T^{\lambda\nu}-x^{\nu}\h T^{\lambda\mu}
\right).
\end{equation}
This form shows that the equilibrium state in Minkowski spacetime is described by the 10 
constant parameters comprised by $b$ and $\varpi$, that is as many as the generators of its maximal 
symmetry group, the Poincar\'e group. 

The familiar form of the equilibrium density operator can be recovered 
as a special case of eq.~(\ref{densop}), with $b$ timelike and $\varpi=0$ in eq.~(\ref{betakilling}):
\begin{equation*}
\h \rho =\frac{1}{Z}\exp[- b \cdot \h P+\zeta \h Q],
\end{equation*}  
which, in the rest frame of $b$, reduces to well known grand-canonical ensamble density
operator. As the above density operator is invariant by translation, we denote this familiar
kind of equilibrium as {\em homogeneous thermodynamic equilibrium}. Another form of global 
equilibrium, which is a special case of eq.~(\ref{densop}) is the pure rotation, with 
$b = (1/T_0,{\bf 0})$ and $\varpi \propto \omega_0/T_0$ such that: 
\begin{equation}\label{rotatop}
\h \rho =\frac{1}{Z} \exp\left[ - \widehat{H}/T_0 + \omega_0 \widehat{J}_z/T_0 \right],
\end{equation}
which has recently raised much attention for fermions \cite{ambrus,chernodub}.

In order to represent a physical fluid at equilibrium, $\beta$ must be a timelike vector. However, 
the equilibrium $\beta$ vector field in eq.~(\ref{betakilling}) is timelike everywhere 
only if $\varpi=0$, corresponding to the homogeneous equilibrium with constant temperature, as we 
have seen. Indeed, if $\varpi\neq 0$ there are spacetime regions where $\beta$ is spacelike 
or lightlike, like in the rotating case (\ref{rotatop}). Nevertheless, this does not undermine 
the possibility to calculate the mean value of local operators in the regions where $\beta$ is timelike. 
The idea is to expand the exponent in the density operator (\ref{densop}) from the same point 
$x$ of a local operator $\h O(x)$, whose mean value is to be calculated, in powers of the 
supposedly small thermal vorticity $\varpi$. Thereby, the zero-order term is the same mean value 
as at homogeneous global equilibrium with four-temperature equal to the $\beta$ field in $x$ 
and corrections arise form a power series in $\varpi$. This method was presented and applied in 
ref.~\cite{Becattini:2015nva} for local operators of the free scalar field and it will be 
outlined in the next section.

\section{Mean value of a local operator}

The mean value of a local operator $\h O(x) $ is defined as: 
\begin{equation}\label{media1}
\media {\h O(x)}=\frac{1}{Z}\tr ( \h \rho\, \h O(x))_{\rm ren}, 
\end{equation}
where the subscript indicates the need of a renormalization procedure; for free fields,
renormalization involves the subtraction of the vacuum expectation value, or, tantamount
the use of normal ordered operators. At global equilibrium, the density operator is given
by eq.~(\ref{densop}) and we can take advantage of this transformation rule for the angular momentum
operator:
\begin{equation}\label{jshift}
\h J^{\mu\nu}_x \equiv \int_\Sigma \di \Sigma_{\lambda} 
\left[ (y^\mu -x^{\mu})\h T^{\lambda\nu}(y)-(y^\nu - x^{\nu})\h T^{\lambda\mu}(y)\right]=
\h J^{\mu\nu}-x^{\mu}\h P^{\nu}+x^{\nu}\h P^{\mu} = \h {\sf T}(x) \h J^{\mu\nu} \h {\sf T}(x)^{-1},
\end{equation}
where $\h {\sf T}(x) = \exp[\ii x \cdot \h P]$ is the translation operator, to rewrite it as:
\begin{equation}\label{densop2}
 \h \rho =\frac{1}{Z} \exp\left[ -\beta_\mu(x) \h P^\mu +\frac{1}{2}\varpi_{\mu\nu}\h J^{\mu\nu}_x 
 +\zeta \h Q\right],
\end{equation}
where $x$ is the point where the operator $\h O(x)$ is to be reckoned and the eq.~(\ref{betakilling}) 
has been used. Thereby, the mean value (\ref{media1}) turns into:
\begin{equation} \label{media2}
 \media {\h O(x)}=\frac{1}{Z}\tr \left[\exp\left(-\beta(x)\cdot \h P+\frac{1}{2}\varpi:\h J_x+
 \zeta\h Q\right)\h O(x)\right]. 
\end{equation}

If $\varpi \ll 1$ (it is adimensional) one can expand the mean value (\ref{media2}) in powers of 
$\varpi$, the leading order being its homogeneous equilibrium value (i.e. the grand-canonical ensemble 
average) with a constant four-temperature $\beta$ equal to the four-temperature field in the point $x$:
\begin{equation}\label{leadorder}
\media{\h O(x)}_{\beta(x)}=\frac{\tr\left[ \exp\left(-\beta(x)\cdot \h P+\zeta \h Q\right)\h O(x)\right]}{\tr\left[ \exp\left(-\beta(x)\cdot \h P+\zeta \h Q\right)\right]}.
\end{equation}
In general, the coefficients of the expansion of (\ref{media2}) in $\varpi$ can be expressed
in terms of correlators at the homogeneous thermodynamic equilibrium. In this work, we will 
use the well-known expansion formula of the exponential of a sum of two non commuting operators 
$\h A+\h B$:
\begin{equation}\label{espansione t ordinata}
 \e^{\h A+\h B}=\e^{\h A} \left[1+\sum_{n=1}^\infty \int_0^1\di \lambda_1 \int_0^{\lambda_1}\di \lambda_2 
 \cdots\int_{0}^{\lambda_{n-1}} \di \lambda_n\h B(\lambda_1)\h B(\lambda_2)\cdots\h B(\lambda_n) \right], 
\end{equation}
where $B(\lambda_1)$ is defined as: 
\begin{equation*}
\h B(\lambda)=\E^{-\lambda \h A}\h B\,\e^{\lambda \h A}.
\end{equation*}
For the mean value in eq.~(\ref{media2}) $\h A=-\beta(x)\cdot \h P+\zeta \h Q $ and 
the small term is $\h B=\frac{1}{2} \varpi:\h J_x$. Therefore, by using the expansion (\ref{espansione 
t ordinata}), we can write:
\begin{equation}
\label{espt1}
 \e^{-\beta(x)\cdot \h P+\zeta \h Q+\frac{1}{2} \varpi:\h J}=\e^{-\beta(x)\cdot \h P+\zeta \h Q}
 \left[1+ \sum_{n=1}^\infty\frac{\varpi^n}{2^n} \int_0^{1}\di \lambda_1 \int_0^{\lambda_1}\di \lambda_2 
  \cdots \int_{0}^{\lambda_{n-1}} \di \lambda_n \h J_{x-\ii \beta \lambda_1}\h J_{x-\ii \beta \lambda_2}
  \cdots \h J_{x-\ii \beta \lambda_n} \right],
\end{equation}
where $\varpi^n$ must be understood as a product of $n$ tensors $\varpi^{\mu\nu}$ with indices fully 
contracted with the coupled angular momentum operators and $\h J_{x-\ii \lambda \beta}$ stands for:
\begin{equation*}
 \h{J}^{\mu\nu}_{x - \ii \lambda \beta}=\e^{\lambda (\beta\cdot \h P-\zeta \h Q)}\h J^{\mu\nu}_x
  \e^{-\lambda (\beta\cdot \h P-\zeta \h Q)} = \e^{\lambda (\beta\cdot \h P-\zeta \h Q)}
   \h J^{\mu\nu}_x \e^{-\lambda (\beta\cdot \h P-\zeta \h Q)} = \h {\sf T}(x - \ii \lambda \beta)
    \h J^{\mu\nu} \h {\sf T}(x - \ii \lambda \beta)^{-1}.
\end{equation*}
This expansion can be written in a more compact form introducing the ${\rm T}_\lambda$-ordered product, where 
the time path is defined on the imaginary direction $ \ii \beta$: 
\begin{equation*}
 {\rm T}_\lambda \big(\h{O}_1(\lambda_1) \h{O}_2(\lambda_2) \cdots \h{O}_N(\lambda_N)\big)
    \equiv \h{O}_{p_1}(\lambda_{p_1}) \h{O}_{p_2}(\lambda_{p_2}) \cdots \h{O}_{p_N}(\lambda_{p_N})
\end{equation*}
with $p$ the permutation that orders $\lambda$ by value:
\begin{align*}
 &p \mathrel{:} \{1, 2, \dots, N\} \to \{1, 2, \dots, N\}\\
 &\lambda_{p_1} \leq \lambda_{p_2} \leq \cdots \leq \lambda_{p_N}.
\end{align*}
Using the above definition and changing the integration limits accordingly, the equation (\ref{espt1}) 
becomes:
\begin{equation*}
\begin{split}
\e^{-\beta(x)\cdot \h P+\zeta \h Q+\frac{1}{2} \varpi:\h J(x)}=&\e^{-\beta(x)\cdot \h P+\zeta \h Q}
\left[1+ \sum_{n=1}^\infty\frac{\varpi^n}{2^nn!} \int_0^{1}\di \lambda_1\di \lambda_2 \cdots 
\di \lambda_n {\rm T}_\lambda \left(\h J_{x-\ii \lambda_1 \beta}\h J_{x-\ii \lambda_2 \beta}\cdots 
 \h J_{x-\ii \lambda_n \beta}\right) \right]\\
= & \e^{-\beta(x)\cdot \h P+\zeta \h Q}{\rm T}_\lambda \left[
\exp\left( \frac{\varpi_{\mu\nu}}{2}\int^{1}_0\di\lambda \h J^{\mu\nu}_{x-\ii \lambda \beta}\right)
\right],
\end{split}
\end{equation*}
where - as usual - the T-ordered exponential is a shorthand notation of the Taylor expansion. 
The mean value of an operator~(\ref{media2}) can now be calculated as a power series of $\varpi$ 
and a cumulant expansion is obtained, namely:  
\begin{equation*}
\media {\h O(x)}=\sum_{N=0}^\infty\frac{\varpi^N}{2^NN!}\int_0^1\di \lambda_1\di \lambda_2\cdots\di \lambda_N\media{
{\rm T}_\lambda\left(\h J_{x-\ii \lambda_1\beta}\h J_{x-\ii\lambda_2\beta}\cdots \h J_{x-\ii \lambda_n\beta} \h O(x)\right)
}_{\beta(x),c},
\end{equation*}
where the correlators are defined respect to the density operator in (\ref{leadorder}), 
$\media{\cdots}_{\beta(x)}=Z^{-1}\tr[e^{-\beta(x)\cdot\h P+\zeta\h Q}\cdots]$ and the subscript 
$c$ means that only the connected correlators are involved, namely:
\begin{equation*}
\begin{split}
\media{\h O}_{c}&=\media{\h O}\\
\media{\h J \h O}_{c}&= \media{\h J\h O}-\media{\h J}\media{\h O}\\
\media{\h J_1\h J_2\h O}_{c}&= \media{\h J_1\h J_2\h O}-\media{\h J_1}\media{\h J_2 \h O}_{c}-
\media{\h J_2}\media{\h J_1 \h O}_{c}-\media{\h O}\media{\h J_1 \h J_2}_{c}-\media{J_1}\media{J_2}\media{\h O}\\
&\vdots
\end{split}
\end{equation*}
So, for instance, at the second order in $\varpi$:
\begin{equation*}
\begin{split}
\media {\h O(x)}=&\media{\h O(x)}_{\beta(x)}+\frac{\varpi_{\mu\nu}}{2}\int_0^1\di \lambda \media{
{\rm T}_\lambda \left(\h J^{\mu\nu}_{x-\ii \lambda\beta}\h O(x)\right)}_{\beta(x),c}
 \\& +\frac{\varpi_{\mu\nu}\varpi_{\rho\sigma}}{8} \int_0^1\di \lambda_1\di \lambda_2 \media{
{\rm T}_\lambda\left(\h J^{\mu\nu}_{x-\ii \lambda_1\beta}\h J^{\rho\sigma}_{x-\ii \lambda_2\beta}\h O(x)\right)
}_{\beta(x),c}+\mathcal{O}(\varpi^3). 
\end{split}
\end{equation*}
This expression can be further worked out by using the local four velocity $u^\mu$ in
eq.~(\ref{fourv}), the inverse proper temperature $|\beta(x)| = 1/T = \sqrt{\beta^2}$ and
changing the integration variable to $\tau = |\beta| \lambda$:
\begin{equation}
\begin{split}\label{tauexp}
\media {\h O(x)}=&\media{\h O(x)}_{\beta(x)}+\frac{\varpi_{\mu\nu}}{2|\beta|}\int_0^{|\beta|}
\di \tau \media{{\rm T}_\tau \left(\h J^{\mu\nu}_{x-\ii \tau u}\h O(x)\right)}_{\beta(x),c}
\\&
+\frac{\varpi_{\mu\nu}\varpi_{\rho\sigma}}{8|\beta|^2}\int_0^{|\beta|}\di \tau_1\di \tau_2
\media{{\rm T}_\tau\left(\h J^{\mu\nu}_{x-\ii \tau_1 u}\h J^{\rho\sigma}_{x-\ii \tau_2 u}\h O(x)\right)
}_{\beta(x),c}+\mathcal{O}(\varpi^3). 
\end{split}
\end{equation}
Since the density operator is the homogeneous one, defined in (\ref{leadorder}), which
is invariant by translation, one can write:
\begin{equation*}
\media{\h O_1(x_1)\h O_1(x_2)\cdots \h O_n (x_n)}_{\beta(x)}=
\media{\h O_1(x_1-x_n)\h O_2(x_2-x_n)\cdots \h O_n (0)}_{\beta(x)},
\end{equation*}
whence in the (\ref{tauexp}):
\begin{equation}
\begin{split}\label{tauexp2}
\media {\h O(x)}=&\media{\h O(0)}_{\beta(x)}+\frac{\varpi_{\mu\nu}}{2|\beta|}\int_0^{|\beta|}\di \tau 
\media{{\rm T}_\tau\left(\h J^{\mu\nu}_{-\ii \tau u}\h O(0)\right)}_{\beta(x),c}
\\& +\frac{\varpi_{\mu\nu}\varpi_{\rho\sigma}}{8|\beta|^2} \int_0^{|\beta|}\di \tau_1\di \tau_2
\media{{\rm T}_\tau\left(\h J^{\mu\nu}_{-\ii \tau_1 u} \h J^{\rho\sigma}_{-\ii \tau_2 u}\h O(0)\right)
}_{\beta(x),c}+\mathcal{O}(\varpi^3).
\end{split}
\end{equation}
This expression makes it apparent that the $x$-dependence of the various terms
is determined only by the four-temperature vector.

\section{Acceleration and vorticity decomposition}\label{accvort}

To proceed, it is useful to decompose the thermal vorticity $\varpi$ by using the four-velocity $u$
in the same fashion as the electromagnetic field tensor is decomposed into comoving electric and
magnetic field. Specifically:
\begin{equation}\label{vortdec}
\varpi_{\mu\nu}=\epsilon_{\mu\nu\rho\sigma}w^\rho u^\sigma+\alpha_\mu u_\nu - \alpha_\nu u_\mu
\end{equation}
where $\alpha$ and $w$ are two spacelike vectors defined as:
\begin{equation}\label{alphaw}
\begin{split}
  \alpha_\mu = \varpi_{\mu\nu} u^\nu,
  \quad w_\mu=-\f{1}{2} \epsilon_{\mu\nu\rho\sigma}\varpi^{\nu\rho}u^\sigma.
\end{split}
\end{equation}
The physical meaning of $\alpha$ and $w$, at the global equilibrium, is that of the acceleration
field and vorticity vector field divided by the proper temperature. To prove this, one has just to show,
by double contracting eq.~(\ref{killing}) with $\beta$, and using the eq.~(\ref{fourv}), that the 
proper temperature does not change along the flow:
$$
u_\mu \partial^\mu {\beta^2} = 0,
$$
so that:
\begin{equation*}
 \alpha^\mu =\varpi^\mu_{\phantom{\mu}\nu} u^\nu= u^\nu \partial_\nu \beta^\mu  
 = \sqrt{\beta^2} u^\nu \partial_\nu u^\mu=\frac{1}{T}a^\mu.
\end{equation*}
Similarly, for $w$:
\begin{equation*}
w_\mu=-\f{1}{2} \epsilon_{\mu\nu\rho\sigma}\varpi^{\nu\rho}u^\sigma=-
\f{1}{2} \epsilon_{\mu\nu\rho\sigma}\partial^\rho\beta^\nu  u^\sigma
=-\f{1}{2} \sqrt{\beta^2}\epsilon_{\mu\nu\rho\sigma}\partial^\rho u^\nu  u^\sigma=\frac{1}{ T}\omega_\mu
\end{equation*}
being $\omega_\mu=-\f{1}{2} \sqrt{\beta^2}\epsilon_{\mu\nu\rho\sigma}\partial^\nu u^\rho  u^\sigma$  
the local vorticity vector. It is also useful to define a fourth four-vector:
\begin{equation} \label{gamma}
 \gamma_\mu=(\alpha\cdot\varpi)^\lambda\Delta_{\lambda\mu}=\epsilon_{\mu\nu\rho\sigma} 
  w^\nu \alpha^\rho u^\sigma,
\end{equation}
where $\Delta_{\mu\nu}=g_{\mu\nu} -u_\mu u_\nu$ is the transverse projector to the four-velocity.
The four-vector $\gamma$ is non-vanishing if $\alpha$ and $w$ are linearly independent and is,
by construction, orthogonal to the other three four-vectors. More relations involving the derivatives
of these vectors can be found in the Appendix \ref{appb}.

In the local rest frame, the above four-vectors can be expressed as: 
\begin{equation*}
 \alpha=\left( 0,\f{\bm{a}}{T} \right),\quad w=\left( 0,\f{\bm{\omega}}{T}\right),\quad 
 \gamma=\left( 0,\f{\bm{a}\wedge\bm{\omega}}{T} \right),
\end{equation*}
where $\bm{a}$ is the acceleration seen by the local rest frame and $\bm{\omega}$ the angular
velocity. Hence, restoring the physical constants:
\begin{equation}\label{vectmagn}
 |\alpha|=\frac{\hbar |\bm{a}|}{c\, k_B\, T},\quad |w|=\frac{\hbar |\bm{\omega}|}{k_B\, T},\quad
 |\gamma|=\frac{\hbar^2 |\bm{a}\times \bm{\omega}|}{c\, k_B^2\, T^2},
\end{equation}
which shows the quantum nature of the adimensional parameters $\alpha,w,\gamma$.

The generators of the Lorentz group $\h J$ can be similarly decomposed into local boosts $\h K$ 
and local angular momenta $\h J$ by using the four velocity $u_\mu$:
\begin{equation*}
\h J^{\mu\nu}=u^\mu \h K^\nu-u^\nu\h K^\mu-u_\rho\epsilon^{\rho\mu\nu\sigma}\h J_\sigma,
\end{equation*}
where the operators $\h K$ and $\h J$ are defined as:
\begin{equation*}
 \h K^\mu=u_\lambda \h J^{\lambda \mu} \quad \quad \h J^\mu = - \frac{1}{2}\epsilon^{\mu\nu\rho\sigma} 
 \h J_{\nu\rho} u_\sigma\, .
\end{equation*}
Therefore, by using the above decomposition and the eq.~(\ref{vortdec}) one can write:
\begin{equation}\label{contract}
\begin{split}
\varpi_{\mu\nu}\h J^{\mu\nu}&=-2\alpha^\rho \h K_{\rho}-2w^\rho\h J_\rho,\\
\varpi_{\mu\nu}\varpi_{\rho\sigma}\h J^{\mu\nu} \h J^{\rho \sigma}&=
2\alpha^{\mu}\alpha^{\nu}\{\h K_\mu,\h K_\nu\}+2w^{\mu}w^{\nu}\{\h J_\mu,\h J_\nu\}
+4\alpha^{\mu}w^{\nu}\{\h K_\mu,\h J_\nu\}, 
\end{split}
\end{equation}
where $\{\cdots,\cdots\}$ is the anti-commutator. Plugging the eqs.~(\ref{contract}) into the 
(\ref{tauexp2}):
\begin{equation}
\begin{split}\label{tauexp3}
&\media {\h O(x)}=\media{\h O(0)}_{\beta(x)}-\frac{\alpha_\rho}{|\beta|}\int_0^{|\beta|}\di \tau \media{
{\rm T}_\tau\left(\h K^{\rho}_{-\ii \tau u} \h O(0)\right)}_{\beta(x),c}-\frac{w_\rho}{|\beta|}
\int_0^{|\beta|}\di \tau \media{{\rm T}_\tau \left(\h J^{\rho}_{-\ii \tau u} \h O(0)\right)}_{\beta(x),c}\\
&+\frac{\alpha_\rho\alpha_\sigma}{2|\beta|^2}\int_0^{|\beta|}\di \tau_1\di \tau_2
\media{{\rm T}_\tau \left(\h K^{\rho}_{-\ii \tau_1 u } \h K^{\sigma}_{- \ii \tau_2 u} \h O(0)\right)}_{\beta(x),c}
+\frac{w_\rho w_\sigma}{2|\beta|^2}\int_0^{|\beta|}\di \tau_1\di \tau_2 \media{{\rm T}_\tau
\left(\h J^{\rho}_{-\ii \tau_1 u } \h J^{\sigma}_{-\ii \tau_2 u} \h O(0)\right)}_{\beta(x),c}
\\
&+\frac{\alpha_\rho w_\sigma}{2|\beta|^2}\int_0^{|\beta|}\di \tau_1\di \tau_2
\media{{\rm T}_\tau \left(\{\h K^\rho_{-\ii \tau_1 u},\h J^{\sigma}_{-\ii \tau_2 u}\}\h O(0)\right)}_{\beta(x),c}
+\mathcal{O}(\varpi^3).
\end{split}
\end{equation}
Now, advantage can be taken of the transformation properties under rotation, reflection and time reversal
of the boost and angular momentum operators to classify all corrections to the mean value of any operator
at thermodynamic equilibrium. 

\section{The stress-energy tensor}

The mean value of the energy momentum tensor $\h T^{\mu\nu}$ receives contributions only 
from second order terms in acceleration and rotation in eq.~(\ref{tauexp3}) because the 
first order terms vanish due to reflection and time reversal symmetry. Thus, the expansion 
(\ref{tauexp3}) becomes:
\begin{equation}
\begin{split}\label{set1}
T^{\mu\nu}(x) = \media {\h T^{\mu\nu}(x)}_{\beta(x)}&+\frac{\alpha_\rho\alpha_\sigma}
{2|\beta|^2}\int_0^{|\beta|}\di \tau_1\di \tau_2 \media{{\rm T}_\tau
\left(\h K^{\rho}_{-\ii \tau_1 u} \h K^{\sigma}_{-\ii \tau_2 u}\h T^{\mu\nu}(0)\right)}_{\beta(x),c}
\\&+\frac{w_\rho w_\sigma}{2|\beta|^2}\int_0^{|\beta|}\di \tau_1\di \tau_2 \media{{\rm T}_\tau
\left(\h J^{\rho}_{-\ii \tau_1 u} \h J^{\sigma}_{-\ii \tau_2 u}\h T^{\mu\nu}(0)\right)}_{\beta(x),c}
\\
&+\frac{\alpha_\rho w_\sigma}{2|\beta|^2}\int_0^{|\beta|}\di \tau_1\di \tau_2 \media{{\rm T}_\tau
\left(\{\h K^{\rho}_{-\ii \tau_1 u},\h J^{\sigma}_{-\ii \tau_2 u} \} \h T^{\mu\nu}(0)\right)}_{\beta(x),c}
+\mathcal{O}(\varpi^3),
\end{split}
\end{equation}
where, according to eq.~(\ref{leadorder}), the mean value of the stress-energy tensor at 
homogeneous thermodynamic equilibrium $\media {\h T^{\mu\nu}(x)}_{\beta(x)}$ coincides with 
its ideal form (\ref{tmunuideal}):
$$
\media {\h T^{\mu\nu}(x)}_{\beta(x)} = \left[\rho\left(\beta^2(x),\mu(x)\right)+p\left(\beta^2(x),\mu(x)\right)\right] u^\mu(x) u^\nu(x) -
  p\left(\beta^2(x),\mu(x)\right) g^{\mu\nu}.
$$
Note that in the above equation we have spelled out all $x$ dependencies and that $\rho,p$ 
are the thermodynamic equilibrium functions of temperature and chemical potential usually 
obtained in thermodynamics.

One can now decompose the above correlators into irreducible tensors under rotation. By taking 
advantage of the rotational invariance of the homogeneous equilibrium density operator in
eq.~(\ref{leadorder}), the number of actual coefficients in eq.~(\ref{set1}) can be reduced, 
and it can be shown \cite{Becattini:2015nva} that:
\begin{equation*}\label{set2}
\begin{split}
\frac{\alpha_\rho\alpha_\sigma}{2|\beta|^2}\int_0^{|\beta|}\di \tau_1\di \tau_2
\media{{\rm T}_\tau \left(\h K^{\rho}_{-\ii \tau_1 u} \h K^{\sigma}_{-\ii \tau_2 u} 
\h T^{\mu\nu}(0)\right)}_{\beta(x),c} &= -\alpha^2 U_\alpha u^\mu u^\nu + \alpha^2 D_\alpha 
\Delta^{\mu\nu} + A \alpha^\mu \alpha^\nu, \\
\frac{w_\rho w_\sigma}{2|\beta|^2}\int_0^{|\beta|}\di \tau_1\di \tau_2 \media{{\rm T}_\tau
\left(\h J^{\rho}_{-\ii \tau_1 u } \h J^{\sigma}_{-\ii \tau_2 u} \h T^{\mu\nu}(0)\right)}_{\beta(x),c}
&=-w^2 U_w u^\mu u^\nu + w^2 D_w \Delta^{\mu\nu}+W w^\mu w^\nu ,\\
\frac{\alpha_\rho w_\sigma}{2|\beta|^2}\int_0^{|\beta|}\di \tau_1\di \tau_2
\media{{\rm T}_\tau \left(\{\h K^{\rho}_{-\ii \tau_1 u} \h J^{\sigma}_{-\ii \tau_2 u} \}
 \h T^{\mu\nu}(0)\right)}_{\beta(x),c} &=G \left( u^\mu\gamma^\nu + u^\nu\gamma^\mu\right),
\end{split}
\end{equation*}
with $\gamma$ as in eq.~(\ref{gamma}). Thereby, the general expression of the 
stress-energy tensor at the second order in thermal vorticity reads \cite{Becattini:2015nva}:
\begin{equation}\label{setgen}
\media{T^{\mu\nu}}=(\rho-\alpha^2 U_\alpha -w^2 U_w)u^\mu u^\nu-(p-\alpha^2D_\alpha-w^2D_w)
\Delta^{\mu\nu}+A\alpha^\mu\alpha^\nu+Ww^\mu w^\nu+
G(u^\mu\gamma^\nu+u^\nu\gamma^\mu)+\mathcal{O}(\varpi^3).
\end{equation}

The coefficients in eq.~(\ref{setgen}) are Lorentz scalars depending only on the magnitude 
of the local inverse four temperature $|\beta|$, that is the proper temperature $T$, and 
the proper chemical potential $\mu$, so they can be calculated in any frame.
The rest frame is the most convenient choice because $\beta^\mu=(1/T,{\bm 0})$ and the 
homogeneous equilibrium density operator takes on the familiar grand-canonical form. We denote 
the mean values in rest frame with a subscript $T$, that is:
$$
 \media{\cdots}_{T}=\frac{\tr\left[\exp\left(-\frac{\h H}{T}+\frac{\mu}{T}\h Q\right)\cdots\right]}{\tr\left[\exp\left(-\frac{\h H}{T}+\frac{\mu}{T}\h Q\right)\right]}.
$$
Hence, the coefficients in (\ref{setgen}) are found as specific combinations 
\cite{Becattini:2015nva} of thermal connected correlators:
\begin{equation}\label{coefficienti}
\begin{split}
U_\alpha=&\frac{1}{2|\beta|^2}\int_0^{|\beta|}\di \tau_1\di \tau_2
\media{{\rm T}_\tau\left(\h K^{3}_{-\ii \tau_1} \h K^{3}_{-\ii \tau_2 }
\h T^{00}(0)\right)}_{T,c}
,\quad 
U_w=\frac{1}{2|\beta|^2}\int_0^{|\beta|}\di \tau_1\di \tau_2 \media{{\rm T}_\tau
\left(\h J^{3}_{-\ii \tau_1} \h J^{3}_{-\ii \tau_2}\h T^{00}(0)\right)}_{T,c},\\
D_\alpha=&\frac{1}{2|\beta|^2}\int_0^{|\beta|}\di \tau_1\di \tau_2
\media{{\rm T}_\tau \left(\h K^{3}_{-\ii \tau_1} \h K^{3}_{- \ii \tau_2} \h T^{11}(0)\right)}_{T,c}
-\frac{1}{3|\beta|^2}\int_0^{|\beta|}\di \tau_1\di \tau_2 \media{{\rm T}_\tau
\left(\h K^{1}_{-\ii \tau_1} \h K^{2}_{-\ii \tau_2} \h T^{12}(0)\right)}_{T,c},\\
D_w=&\frac{1}{2|\beta|^2}\int_0^{|\beta|}\di \tau_1\di \tau_2 \media{{\rm T}_\tau
\left(\h J^{3}_{-\ii \tau_1} \h J^{3}_{-\ii \tau_2} \h T^{11}(0)\right)}_{T,c}
-\frac{1}{3|\beta|^2}\int_0^{|\beta|}\di \tau_1\di \tau_2 \media{{\rm T}_\tau
\left(\h J^{1}_{-\ii \tau_1} \h J^{2}_{-\ii \tau_2} \h T^{12}(0)\right)}_{T,c},
\\
A=&\frac{1}{|\beta|^2}\int_0^{|\beta|}\di \tau_1\di \tau_2 \media{{\rm T}_\tau
\left(\h K^{1}_{-\ii \tau_1} \h K^{2}_{-\ii \tau_2} \h T^{12}(0)\right)}_{T,c},
\quad
W=\frac{1}{|\beta|^2}\int_0^{|\beta|}\di \tau_1\di \tau_2 \media{{\rm T}_\tau
\left(\h J^{1}_{-\ii \tau_1} \h J^{2}_{-\ii \tau_2} \h T^{12}(0)\right)}_{T,c}
\\
G=&-\frac{1}{2|\beta|^2}\int_0^{|\beta|}\di \tau_1\di \tau_2 \media{{\rm T}_\tau
\left(\{\h K^{1}_{-\ii \tau_1},\h J^{2}_{-\ii \tau_2} \}\h T^{03}(0)\right)}_{T,c}\, .
\end{split}
\end{equation}

The correlation functions in (\ref{coefficienti}) can be expressed as Euclidean three-point 
functions of the stress-energy tensor and can be calculated with the imaginary time formalism.
The shifted boost and angular momentum generators can be written as (see eq.~(\ref{jshift})):
\begin{equation*}
  \h J^{\mu\nu}_{-\ii \tau}= \h {\sf T}\left((-\ii \tau, {\bf 0})\right) \h J^{\mu\nu} \h {\sf T}^{-1}\left((-\ii \tau, {\bf 0})\right).
\end{equation*}
So one can expand the integral expressions of the generators of the Lorentz group and 
write the basic structure appearing in all coefficients in (\ref{coefficienti}) as:
\begin{equation}
\begin{split}\label{C-stress}
C^{\alpha\beta|\gamma\rho|\mu\nu|ij}=\frac{1}{|\beta|^2}\int_0^{|\beta|}\di \tau_1\di \tau_2
\int\di^3x \,\, \di^3y \media{{\rm T}_\tau \left(\h T^{\alpha\beta}(\tau_1 ,{\bm x}) 
\h T^{\gamma\rho}(\tau_2 ,{\bm y}) \h T^{\mu\nu}(0)\right)}_{T,c} x^i y^j
\end{split}
\end{equation}
where 
\be\label{imset}
 \h T^{\mu\nu} (\tau, {\bf x}) = \h {\sf T}\left((-\ii \tau, {\bf 0})\right) \h T^{\mu\nu}(0,{\bf x}) 
 \h {\sf T}^{-1}\left((-\ii \tau, {\bf 0})\right)
\ee 
is the imaginary time evolved stress-energy tensor operator.
By using the eq.~(\ref{C-stress}), the eqs.~(\ref{coefficienti}) can be rewritten in terms of 
these auxiliary quantities:
\begin{equation}\label{coefficientiC}
\begin{split}
U_\alpha=&\frac12C^{00|00|00|33}
,\quad 
U_w=\frac12\left( C^{01|01|00|22}-C^{01|02|00|21}-C^{02|01|00|12}+C^{02|02|00|11}\right),\\
D_\alpha=&\frac12C^{00|00|11|33}-\frac13 C^{00|00|12|12}
,\\
D_w=& 
\frac12 \left(
C^{01|01|11|22}-C^{01|02|11|21}-C^{02|01|11|12}+C^{02|02|11|11}
\right)
\\&-\frac13 \left(
C^{02|03|12|31}-C^{03|03|12|21}-C^{02|01|12|33}+C^{03|01|12|23}
\right),
\\
A=&\, C^{00|00|12|12},\quad
W=C^{02|03|12|31}-C^{03|03|12|21}-C^{02|01|12|33}+C^{03|01|12|23},
\\
G=&\frac12\left(C^{00|03|03|11}-C^{00|01|03|13}+C^{03|00|03|11}-C^{01|00|03|31}\right).
\end{split}
\end{equation}

The above coefficients are not all independent, in fact there are relations between them
stemming from the conservation equation: 
\begin{equation*}
\partial_\mu\media{\h T^{\mu\nu}}= \partial_{\mu}\frac{1}{Z}\tr (\h \rho \, \h T^{\mu\nu})
=\media{\partial_\mu \h T^{\mu\nu}}=0,
\end{equation*}
which states that the mean value of $\h T^{\mu\nu}$ is conserved if the corresponding 
operator is conserved and the density operator $\h\rho$ is time-independent. The relation
between the above second-order coefficients can be obtained in a functional approach \cite{Bhattacharyya:2012nq,kovtun,Banerjee:2012iz} by taking advntage of the invariance
of the generating functional by diffeomorfisms; in the present operatorial approach, one 
has to enforce the continuity equation of the stress-energy tensor, which leads to these 
equalities (see Appendix \ref{appb}):
\begin{equation}\label{coeffrel}
\begin{split}
 U_\alpha&=-|\beta|\f{\partial}{\partial|\beta|}\big(D_\alpha+A\big)-\big(D_\alpha+A\big),\\
 U_w&=-|\beta|\f{\partial}{\partial|\beta|}\big(D_w+W\big)-D_w+2A-3W,\\
 2G&=2\big(D_\alpha+D_w\big)+A+|\beta|\f{\partial}{\partial|\beta|}W+3W,
\end{split}
\end{equation}
where all derivatives are to be taken with fixed $\zeta=\mu |\beta|$.
Thus, only $D_\alpha,D_w,A,W$ are actually independent, while $G,U_\alpha,U_w$ can be
obtained directly from the eqs.~(\ref{coeffrel}). Nevertheless, in this work, we have calculated 
all the coefficients by using the eqs.~(\ref{coefficienti}) and have used the relations in
eq.~(\ref{coeffrel}) as a consistency check.

\subsection{Free complex scalar field}

We now calculate the coefficients (\ref{coefficienti}) for a complex scalar field at finite 
temperature and chemical potential, using the imaginary time formalism. In this case we have:
\begin{equation}\label{sset}
\h T_{\mu\nu}= \partial_\mu\wpsi^\dagger \partial_\nu \wpsi+\partial_\nu\wpsi^\dagger \partial_\mu 
\wpsi-g_{\mu\nu} ( \partial\wpsi^\dagger \cdot \partial \wpsi -m^2 \wpsi^\dagger \wpsi )
- \xi (\partial_\mu \partial_\nu - g_{\mu\nu}\square ) \wpsi^\dagger \wpsi\\
\end{equation}
%
where $\xi=0$ corresponds to the \emph{canonical} stress-energy tensor and $\xi=1/6$ to the 
improved stress-energy tensor \cite{callan}.

For finite chemical potential, it is convenient to switch to the Euclidean time formalism 
with the modified hamiltonian $\h H-\mu \h Q$ and write the field evolution as:
\begin{equation*}
 \wpsi(\tau,\vec{x} )= \e^{\tau (\h H- \mu \h Q)} \wpsi(0,\vec{x}) \e^{-\tau (\h H- \mu \h Q)}.
\end{equation*}
Indeed, as the stress-energy tensor commutes with the charge operator, the formulae obtained
in the previous section are unaffected. Defining:
\begin{equation}\label{symbols}
P^\pm=(\omega_n\pm\ii\mu, {\bm p} ),\quad \quad X=(\tau ,{\bm x}), 
\quad \quad\SumInt_{P}=\frac{1}{|\beta|}\sum_{n=-\infty}^\infty\int \frac{\di ^3 p }{(2\pi)^3}\, ,
\quad \int_X=\int_0^{|\beta|}\di \tau\int \di^3x 
\end{equation}
and: 
$$
\tilde\delta(P)=\int_X \e^{\ii X\cdot P}=|\beta| (2\pi)^3 \delta_{p_n,0}\,\delta^{(3)}({\bm p})
$$
where $\omega_n=2\pi n /|\beta|$ are the Matsubara frequencies, the propagator in the imaginary
time reads \cite{Kapusta:2006pm,Laine:2016hma}:
\begin{equation}\label{propagator}
\begin{split}
\media{ {\rm T}_\tau \wpsi^\dagger(X)\wpsi(Y)}_T&=\SumInt_{P}\e^{\ii P \cdot (X-Y)-\mu(\tau_x-\tau_y)} 
\frac{1}{(\omega_n+\ii \mu)^2+{\bm p}^2+m^2}\equiv \SumInt_{P}\e^{\ii P^+(X-Y)} \Delta(P^+),\\
\media{ {\rm T}_\tau \wpsi(X)\wpsi^\dagger(Y)}_T&=\SumInt_{P}\e^{\ii P \cdot (X-Y)+\mu(\tau_x-\tau_y)} 
\frac{1}{(\omega_n-\ii \mu)^2+{\bm p}^2+m^2}\equiv \SumInt_{P}\e^{\ii P^-(X-Y)} \Delta(P^-),
\end{split}
\end{equation}
where:
\begin{equation}\label{propden}
 \Delta(P^\pm) = \frac{1}{{P^\pm}^2+m^2}
\end{equation}
and ${P^\pm}^2$ is the Euclidean squared magnitude of the four vector $P^\pm$, implicitly 
defined in eqs.~(\ref{propagator}). The eqs.~(\ref{propagator}) are the building block to evaluate the 
three-point function in eq.~(\ref{C-stress}) because:
\begin{equation}\label{bblock}
\begin{split}
C^{\alpha\beta|\gamma\rho|\mu\nu|ij}=\frac{1}{|\beta|^2} \int_X \int_Y \media{{\rm T}_\tau 
 \h T^{\alpha\beta}(X) \h T^{\gamma\rho}(Y) \h T^{\mu\nu}(0)}_{T,c} x^i y^j
\end{split}
\end{equation}
where $\h T^{\mu\nu}(X)$ is given by (\ref{imset}). The (\ref{bblock}) can be computed using 
the point splitting procedure; first, consider:
\begin{equation}\label{3pf}
\begin{split}
C(X,Y,Z)&\equiv\media{{\rm T}_\tau \h T^{\alpha\beta}(X)\h T^{\gamma\rho}(Y) \h T^{\mu\nu}(Z)}_{T,c} \\
&=\lim_{\substack{X_1,X_2\to X \\ Y_1,Y_2\to Y\\ 
Z_1,Z_2\to Z}}\mathcal{D}^{\alpha\beta}(\partial_{X_1},\partial_{X_2}) \mathcal{D}^{\gamma\rho}
(\partial_{Y_1},\partial_{Y_2}) \mathcal{D}^{\mu\nu}(\partial_{Z_1},\partial_{Z_2})
\media{{\rm T}_\tau\wpsi^\dagger(X_1)\wpsi(X_2)\wpsi^\dagger(Y_1)\wpsi(Y_2)\wpsi^\dagger(Z_1)\wpsi(Z_2)}_{T,c}
\end{split}
\end{equation}
where the form of the differential operators $\mathcal{D}^{\mu\nu}(\partial_{X_1},\partial_{X_2})$ can 
be inferred from the eq.~(\ref{sset}):
\begin{equation}\label{polinomio}
\begin{split}
\mathcal{D}_{\alpha\beta}(\partial_{X_1},\partial_{X_2}) &= (-\ii)^{\delta_{0\alpha}+\delta_{0\beta}}
\left[ (1-2\xi)\left(\partial_{X_1\alpha}\partial_{X_2\beta} + \partial_{X_2\alpha} \partial_{X_1\beta}\right)
-(1-4\xi)\delta_{\alpha\beta} \partial_{X_1} \cdot  \partial_{X_2} -m^2\delta_{\alpha\beta}
\right.\\
&\left.+2\xi( \delta_{\alpha\beta}\square_{X_2E}+\delta_{\alpha\beta}
\square_{X_1E} -\partial_{X_2\alpha}\partial_{X_2\beta}-\partial_{X_1\alpha}\partial_{X_1\beta})\right].
\end{split}
\end{equation}
In the eq.~(\ref{polinomio}), the scalar product has the Euclidean signature, that is $\partial_{X_1} 
\cdot  \partial_{X_2}= \partial_{\tau_1}\partial_{\tau_2}+\sum_i \partial_{x_{i1}}\partial_{x_{i2}}$ and the
D'Alembertian as well: $\square_{XE}=\partial_\tau^2+\sum_i \partial_i^2$. The imaginary unit in front 
of the differential operator $(-\ii)^{\delta_{0\alpha}+\delta_{0\beta}}$ is a consequence of the the Wick 
rotation. The evaluation of the three-point function can be done by employing the standard Wick 
theorem and since only its connected part appears in (\ref{3pf}), only the following 
two terms survive:
\begin{equation*}
\begin{split}
 \media{{\rm T}_\tau\wpsi^\dagger(X_1)\wpsi(X_2)\wpsi^\dagger(Y_1)\wpsi(Y_2)\wpsi^\dagger(Z_1)\wpsi(Z_2)}_{T,c}
 =&\media{{\rm T}_\tau\wpsi^\dagger(X_1)\wpsi(Y_2)} \media{{\rm T}_\tau\wpsi(X_2)\wpsi^\dagger(Z_1)} 
 \media{{\rm T}_\tau\wpsi^\dagger(Y_1)
 \wpsi(Z_2)}\\
 &+ \media{{\rm T}_\tau\wpsi^\dagger(X_1)\wpsi(Z_2)}\media{{\rm T}_\tau\wpsi(X_2)\wpsi^\dagger(Y_1)} \media{{\rm T}_\tau\wpsi(Y_2)\wpsi^\dagger(Z_1)}.
\end{split}
\end{equation*}
Inserting the Fourier decomposition of eq.~(\ref{propagator}), the differential operators in
(\ref{3pf}) give rise to a polynomial in momentum; thereby, the limits can be readily done 
and one obtains:
\begin{equation}\label{Cxyz}
\begin{split}
C(X,Y,Z)=\SumInt_{P,Q,K}\e^{\ii P\cdot(X-Y)+Q\cdot(Y-Z)+K\cdot(X-Z)}&\left[ F_1(P^+,Q^+,K^-) 
\Delta(P^+)\Delta(K^-)\Delta(Q^+) \right.\\
&\left.+F_2(P^-,Q^-,K^+) \Delta(P^-)\Delta(K^+)\Delta(Q^-) \right]
\end{split}
\end{equation}
where $\Delta$ is defined in (\ref{propden}) and the function $F_1$ and $F_2$ are just polynomials 
of momenta:
\begin{equation*}
\begin{split}
F_1(P^+,Q^+,K^-)=& \mathcal{D}^{\alpha\beta}(\ii P^+,\ii K^-) 
\mathcal{D}^{\gamma\rho}(\ii Q^{+},-\ii P^{+}) \mathcal{D}^{\mu\nu}(-\ii K^{-},-\ii Q^+),\\
F_2(P^-,Q^-,K^+)=&\mathcal{D}^{\alpha\beta}(\ii K^+,\ii P^-) 
\mathcal{D}^{\gamma\rho}(-\ii P^{-},\ii Q^{-}) \mathcal{D}^{\mu\nu}(-\ii Q^{-},-\ii K^+) .
\end{split}
\end{equation*}
Now we take the $Z \to 0$ limit in eq.~(\ref{Cxyz}), as prescribed by eq.~(\ref{bblock}) and, 
separating the integration from the sum over frequencies, we get:
\begin{equation}\label{CC}
\begin{split}
C(X,Y,0) = C(\tau_1,{\bm x},\tau_2,{\bm y}) =\int\frac{\di^3 p}{(2\pi)^3}\frac{\di^3 k}{(2\pi)^3}
\frac{\di^3 q}{(2\pi)^3}\e^{-\ii ({\bm p}+{\bm k} )\cdot {\bm x}}\e^{-\ii({\bm q}-{\bm p} )
\cdot {\bm y} }S({\bm p},{\bm q},{\bm k},\tau_1, \tau_2)
\end{split}
\end{equation}
where:
\begin{equation}\label{s3p}
\begin{split}
S({\bm p},{\bm q},{\bm k},\tau_1, \tau_2)=\frac{1}{|\beta|^3}\sum_{p_n,q_n,k_n} &
\frac{\e^{\ii (p_n+\ii\mu)(\tau_1-\tau_2)+\ii (q_n+\ii\mu)\tau_2+\ii (k_n-\ii\mu)\tau_1}F_1(P^+,Q^+,K^-)}{({P^+}^2 +m^2)({Q^+}^2 +m^2)({K^-}^2 +m^2)}\\
&+\frac{\e^{\ii (p_n-\ii\mu)\tau(\tau_1-\tau_2)+\ii (q_n-\ii\mu)-\tau_2+\ii (k_n+\ii\mu)\tau_1}F_2(P^-,Q^-,K^+)}{({P^-}^2 +m^2)({Q^-}^2 +m^2)({K^+}^2 +m^2)}\, .
\end{split}
\end{equation}
The functions $F_1$ and $F_2$ are polynomials, hence analytic and the sum over the frequencies 
can be carried out by using the formula \cite{Landsman:1986uw} :
\begin{equation}\label{sommabosoni}
\begin{split}
\frac{1}{|\beta|}\sum_{\omega_n}\frac{(\omega_n\pm \ii \mu)^k \e^{\ii (\omega_n\pm \ii \mu) \tau}}
{(\omega_n\pm \ii \mu)^2+E^2}&=\frac{1}{2E}\left[(-\ii E)^k\e^{\tau E}(\theta(-\tau)+n_B(E\pm\mu))+
(\ii E)^k\e^{-E\tau}(\theta(\tau)+n_B(E\mp\mu)) \right]
\\
&=\frac{1}{2E}\sum_{s=\pm 1} (-\ii s E)^k\e^{\tau  s E}[\theta(- s \tau)+n_B(E\pm s \mu)]
\end{split}
\end{equation}
where $\theta(\tau)$ is the Heaviside function and $-|\beta| <\tau <|\beta|$; $n_B$ is the 
Bose distribution function:
\begin{equation}\label{bose}
n_{B}(E)=\frac{1}{\e^{|\beta|E}-1}\, .
\end{equation}
The eq.~(\ref{sommabosoni}) is needed to work out the function in eq.~(\ref{s3p}):
\begin{equation*}
\begin{split}
&S({\bm p},{\bm q},{\bm k},\tau_1, \tau_2)=\frac{1}{8E_p E_q E_k }
\sum_{s_1,s_2,s_3=\pm 1}\e^{\tau_1 ( s_1 E_p +s_3 E_k )+\tau_2(s_2 E_q -s1E_q  )}
\left\{ F_1\left(\tilde P(s_1) ,\tilde Q(s_2) ,\tilde K(s_3) \right)\right. \\
& \left.
[\theta(- s_1( \tau_1-\tau_2))+n_{B}(E_p +s_1\mu)]  [\theta(- s_2 )+n_{B}(E_q +s_2\mu)]   
[\theta(- s_3 )+n_{B}(E_k -s_3\mu)] +
\right.\\
&
\left. F_2\left(\tilde P(s_1) ,\tilde Q(s_2) ,\tilde K(s_3) \right)
[\theta(- s_1( \tau_1-\tau_2))+n_{B}(E_p -s_1\mu)]  [\theta(- s_2 )+n_{B}(E_q -s_2\mu)]   
[\theta(- s_3 )+n_{B}(E_k +s_3\mu)]
\right\}
\end{split}
\end{equation*}
where $E_p=\sqrt{{\bm p}^2+m^2}$, and we have defined
\begin{equation}\label{newmom}
\tilde P(s)\equiv(-\ii s E_p,\,{\bm p})
\end{equation}
and similarly for $\tilde Q(s),\,\tilde K(s)$.
%
Note that, after the frequency summation, the arguments of the functions $F_1$ and $F_2$ no longer depend 
on the chemical potential, and that, thanks to the symmetry properties of the polynomials, 
the functions $F_1$ and $F_2$ become indeed the same. Thus, the eq.~(\ref{s3p}) becomes:
\begin{equation}\label{s3pf}
\begin{split}
&S({\bm p},{\bm q},{\bm k},\tau_1, \tau_2)=\frac{1}{8E_p E_q E_k }\sum_{s_1,s_2,s_3=\pm 1}
\e^{\tau_1 ( s_1 E_p +s_3 E_k )+\tau_2(s_2 E_q -s_1E_q  )}F_1\left(\tilde P(s_1) ,\tilde Q(s_2) ,\tilde K(s_3) \right)\\
& \Big\{ [\theta(- s_1( \tau_1-\tau_2))+n_{B}(E_p -s_1\mu)]  [\theta(- s_2 )+n_{B}(E_q -s_2\mu)]   
[\theta(- s_3 )+n_{B}(E_k +s_3\mu)]+(\mu \to -\mu) \Big\}.
\end{split}
\end{equation}
Now we can take advantage of the formula:
\begin{equation*}
\begin{split}
\int \di^3 x \int\di^3 y \,\, \e^{-\ii ({\bm p}+{\bm k} )\cdot {\bm x}}e^{-\ii({\bm q}-{\bm p} )\cdot {\bm x} } 
x^i y^j=-(2\pi)^6 \frac{\partial^2}{\partial k_i\partial q_j}\delta^{(3)}(\bm p +\bm k)\delta^{(3)}(\bm p -\bm q)
\end{split}
\end{equation*}
to integrate over the coordinates $x$ and $y$ in the eq.~(\ref{bblock}) and, by using the eq.~(\ref{CC})
we obtain: 
\begin{equation}\label{finbos}
\begin{split}
C^{\alpha\beta|\gamma\rho|\mu\nu|ij}=\frac{1}{|\beta|^2} \int_X \int_Y \media{{\rm T}_\tau
\h T^{\alpha\beta}(X)\h T^{\gamma\rho}(Y) \h T^{\mu\nu}(0)
}_{T,c} x^i y^j=\frac{-1}{|\beta|^2}\int^{|\beta|}_0\di \tau_1 \di\tau_2\int \frac{\di^3 p}{(2\pi)^3} 
\frac{\partial^2}{\partial k_i\partial q_j}S({\bm p},{\bm q},{\bm k},\tau_1, \tau_2)
\Big|_{\substack{{\bm q}= {\bm p}\\ {\bm k}= -{\bm p} }}\, .
\end{split}
\end{equation}
One can now plug the (\ref{s3pf}) into the (\ref{finbos}), integrate over $\tau_1$ and $\tau_2$ 
so as to express the $C^{\alpha\beta|\gamma\rho|\mu\nu|ij}$ as an integral over momentum of combinations of
derivatives of the Bose distribution function. After setting the appropriate indices in (\ref{finbos})
according to the (\ref{coefficientiC}), the second-order coefficients of the stress-energy tensor
are finally obtained:
\begin{equation}\label{boscoeff}
\begin{split}
U_w&=\frac{(1-4\xi)}{12\pi^2\beta^2}\int_0^\infty \frac{\di p}{E_p}  
\left(n_B''(E_p-\mu)+n_B''(E_p+\mu)\right) p^4,\\
U_\alpha&=\frac{1}{48 \pi^2\beta^2}\int_0^\infty \frac{\di p}{E_p}\left(n_B''(E_p-\mu)+n_B''(E_p+\mu)\right) 
 (p^2+m^2)(m^2+4p^2(1-6\xi)), \\
W&=\frac{(2\xi-1)}{12\pi^2\beta^2}\int_0^\infty \frac{\di p}{E_p}\left(n_B''(E_p-\mu)+n_B''(E_p+\mu)\right) p^4 
,\\
A&= \frac{1}{24\pi^2\beta^2}\int_0^\infty \frac{\di p}{E_p}  \left(n_B''(E_p-\mu)+n_B''(E_p+\mu)\right)  
\left(2p^4(1-6\xi)+p^2m^2(3-12\xi)\right) ,\\
G&= \frac{1}{36\pi^2\beta^2}\int_0^\infty \frac{\di p}{E_p}  \left(n_B''(E_p-\mu)+n_B''(E_p+\mu)\right)  
\left(p^4(1+6\xi)+3p^2m^2\right) ,\\
D_\alpha&= \frac{1}{144\pi^2\beta^2}\int_0^\infty \frac{\di p}{E_p}  \left(n_B''(E_p-\mu)+n_B''(E_p+\mu)\right)
 \left(8p^4(6\xi-1)+3p^2m^2(24\xi-5)\right),\\
D_w&= \frac{\xi}{6\pi^2\beta^2}\int_0^\infty \frac{\di p}{E_p}  \left(n_B''(E_p-\mu)+n_B''(E_p+\mu)\right)p^4,
\end{split}
\end{equation}
where $n_B''$ denotes the second derivative of the Bose distribution function with respect
to the energy. 

It is worth noticing that the above expressions~(\ref{boscoeff}) fulfill the relations~(\ref{coeffrel}).

\subsection{Free Dirac field } 

In this section we calculate the coefficients (\ref{coefficienti}) for the free Dirac field 
at finite temperature and chemical potential. The thermodynamic properties of the Dirac field 
can be deduced from the Euclidean Lagrangian $\h{{\cal L}}_E=-\h{\mathcal{L}}(t=-\ii \tau)$:
\begin{equation}\label{edlag}
\h{{\cal L}}_E=\frac{1}{2}\left[\wbPsi \gamma^0 \partial_\tau \wPsi -\partial_\tau \wbPsi  
\gamma^0\wPsi -\ii \wbPsi \gamma^k \partial_k \wPsi-\ii \partial_k \wbPsi  \gamma^k\wPsi\right]
+m\wbPsi \wPsi.
\end{equation}
It is convenient to write the (\ref{edlag}) with the so-called Euclidean Dirac matrices:
\begin{equation*}
\tilde\gamma_0\equiv\gamma^0,\quad \quad \tilde \gamma_k =-\ii \gamma^k 
\end{equation*}
fulfilling the relations 
$$
\{\tilde \gamma_\mu ,\tilde \gamma_\nu \}=2 \delta_{\mu\nu} \qquad \qquad
\tilde \gamma_\mu^\dagger=\tilde\gamma_\mu
$$
so that the eq.~(\ref{edlag}) can be written:
\begin{equation*}
\h{{\cal L}}_E=\frac{1}{2}\left[\wbPsi\tilde\gamma_\mu\partial_\mu\wPsi-\partial_\mu\wbPsi
\tilde\gamma_\mu\wPsi\right]+m \wbPsi\wPsi .
\end{equation*}
The thermal propagator of the Dirac field reads \footnote{Latin characters $a,b,\cdots$ denote
spinorial indices}\cite{Kapusta:2006pm,Laine:2016hma}:
\begin{equation}\label{dirpropag}
\media{{\rm T}_\tau \wPsi_a(X) \wbPsi_b (Y)}= \SumInt_{\{P\}}\e^{\ii P^+\cdot (X-Y)}
\frac{(-\ii \slashed P^++m)_{ab}}{(P^+)^2+m^2} = \SumInt_{\{P\}}\e^{\ii P^+\cdot (X-Y)}
(-\ii \slashed P^++m)_{ab}\Delta(P^+)
\end{equation}
where $X,Y,P^{\pm}$ are defined in eq.~(\ref{symbols}) and $\Delta(P^\pm)$ in 
eq.~(\ref{propden}); the sum runs over the fermionic Matsubara frequencies 
$\omega_n =2\pi(n+\frac{1}{2})/|\beta|$; $\slashed P^{\pm}=\tilde\gamma_\mu P_\mu^\pm$ is 
the standard contraction between the (Euclidean) Dirac matrices $\tilde\gamma_\mu$ and 
the (Euclidean) four-momenta $P^\pm=(p_n \pm \ii \mu,{\bm p})$. The Euclidean canonical 
stress-energy tensor (see eq. (\ref{imset})) reads:
\begin{equation*}
\h T_{\mu\nu}(X)=\frac{\ii^{\delta_{0\mu}+\delta_{0\nu}}}{2}
\left[\wbPsi(X) \tilde \gamma_\mu \partial_\nu \wPsi(X) -\partial_\nu \wbPsi(X) \tilde \gamma_\mu \wPsi(X)\right]
\end{equation*}
where the $\ii^{\delta_{\nu 0}}$ factor stems from Wick rotation. The Belinfante-symmetrized 
stress-energy tensor is the symmetric part of the canonical one:
\begin{equation*}
\h T_{\mu\nu}(X)=\frac{\ii^{\delta_{0\mu}+\delta_{0\nu}}}{4}\left[\wbPsi(X) \tilde \gamma_\mu 
\partial_\nu \wPsi(X) -\partial_\nu \wbPsi(X) \tilde \gamma_\mu \wPsi(X)+\wbPsi(X) 
\tilde \gamma_\nu \partial_\mu \wPsi(X) -\partial_\mu \wbPsi(X) \tilde \gamma_\nu \wPsi(X)\right],
\end{equation*}
which can be expressed according to the point-splitting procedure as:
\begin{equation*}
\h T_{\mu\nu}(X)= \lim_{X_1,X_2\to X} \mathcal{D}_{\mu\nu} (\partial_{X_1},\partial_{X_2})\wbPsi (X_1)\wPsi (X_2),
\end{equation*}
where:
\begin{equation}\label{fermD}
 \mathcal{D}_{\mu\nu} (\partial_{X_1},\partial_{X_2})=
 \frac{\ii^{\delta_{0\mu}+\delta_{0\nu}}}{4}\left[\tilde \gamma_\mu (\partial_{X_2}-\partial_{X_1})_\nu
 +\tilde\gamma_\nu (\partial_{X_2}-\partial_{X_1})_\mu\right].
\end{equation}

The stress-energy tensor three-point correlation function (\ref{bblock}) needed to extract the various 
coefficients is similar to that in (\ref{3pf}):
\begin{equation*}
\begin{split}
C(X,Y,Z)&=\media{{\rm T}_\tau\h T^{\alpha\beta}(X)\h T^{\gamma\rho}(Y) \h T^{\mu\nu}(Z)}_{T,c} =\\
\lim_{\substack{X_1,X_2\to X \\ Y_1,Y_2\to Y\\ 
Z_1,Z_2\to Z}}&\mathcal{D}^{\alpha\beta}(\partial_{X_1},\partial_{X_2})_{ab} \mathcal{D}^{\gamma\rho}
(\partial_{Y_1},\partial_{Y_2})_{cd} \mathcal{D}^{\mu\nu}(\partial_{Z_1},\partial_{Z_2})_{ef}
 \media{{\rm T}_\tau\wbPsi_a(X_1)\wPsi_b(X_2)\wbPsi_c(Y_1)\wPsi_d(Y_2)\wbPsi_e(Z_1)\wPsi_f(Z_2)}_{T,c}\, .
\end{split}
\end{equation*}
Like for the boson case, thanks to the Wick theorem and the presence of the connected part, only 
two contractions survive:
\begin{equation*}
\begin{split}
  \media{{\rm T}_\tau\wbPsi_a(X_1)\wPsi_b(X_2)\wbPsi_c(Y_1)\wPsi_d(Y_2)\wbPsi_e(Z_1)\wPsi_f(Z_2)}_{T,c}=
  &\media{{\rm T}_\tau\wbPsi_a(X_1)\wPsi_d(Y_2)} \media{{\rm T}_\tau\wPsi_b(X_2)\wbPsi_e(Z_1)} \media{{\rm T}_\tau\wbPsi_c(Y_1)\wPsi_f(Z_2)}\\+&\media{{\rm T}_\tau\wbPsi_a(X_1)\wPsi_f(Z_2)}\media{{\rm T}_\tau\wPsi_b(X_2)\wbPsi_c(Y_1)}
  \media{{\rm T}_\tau\wPsi_d(Y_2)\wbPsi_e(Z_1)}\, .
\end{split}
\end{equation*}
Plugging the expression of the propagators (\ref{dirpropag}), after some algebra and integration
variable manipulation, we get:
\begin{equation*}
\begin{split}
C(X,Y,Z)=(-1)\SumInt_{\{P,Q,K\}}\e^{\ii P\cdot(X-Y)+Q\cdot(Y-Z)+K\cdot(X-Z)}&
\left[ G_1(P^+,Q^+,K^-) \Delta(P^+)\Delta(K^-)\Delta(Q^+) \right.\\ &\left.+G_2(P^-,Q^-,K^+)
\Delta(P^-)\Delta(K^+)\Delta(Q^-) \right]
\end{split}
\end{equation*}
where the functions $G_1$ and $G_2$ are defined as:
\begin{equation*}
\begin{split}
G_1(P^+,Q^+,K^-) &=\tr\left[
\mathcal{D}_{\alpha\beta}(\ii K^-,\ii P^+)(-\ii\slashed P^++m)
\mathcal{D}_{\gamma\rho}(-\ii P^+,\ii Q^+)(-\ii\slashed Q^++m)
\mathcal{D}_{\mu\nu}(-\ii Q^+,-\ii K^-)(\ii\slashed K^-+m)\right],
\\
G_2(P^-,Q^-,K^+)& =\tr\left[
\mathcal{D}_{\alpha\beta}(\ii P^-,\ii K^+)(-\ii\slashed K^++m)\mathcal{D}_{\mu\nu}(-\ii K^+,-\ii Q^-)(\ii\slashed Q^-+m)\mathcal{D}_{\gamma\rho}(\ii Q^-,-\ii P^-)(\ii\slashed P^++m) \right].
\end{split}
\end{equation*}
The trace is to be carried out over spinorial indices by using the Euclidean $\gamma$ 
matrices properties:
 \begin{gather*}
 \tr\left( \tilde\gamma_\mu \tilde\gamma_\nu\right)=4\,\delta_{\mu\nu}, \\
 \tr\left( \tilde\gamma_{\mu_{1}} \dots \tilde\gamma_{\mu_{2n+1}}\right)=0, \\
 \tr\left( \tilde\gamma_k \gamma_\lambda \tilde\gamma_\mu \tilde\gamma_\nu\right)=4\, 
 \delta_{k\lambda}\delta_{\mu\nu}-4\, \delta_{k\mu}\delta_{\lambda\nu}+4\, \delta_{k\nu}\delta_{\lambda\mu}.
\end{gather*}
We finally obtain:
\begin{equation*}
\begin{split}
 &G_1(P,Q,K)=G_2(P,Q,K) = {\cal S}_{\alpha\beta\gamma\rho\mu\nu} \Big\{ \f{1}{2} 
 \left(P_\beta-K_\beta\right) \left(P_\rho+Q_\rho\right) \left(K_\nu-Q_\nu\right)\cdot\\
& \cdot\Big[K_\mu \Big(P_\gamma Q_\alpha+P_\alpha Q_\gamma-\delta_{\alpha\gamma} \left(m^2+P\cdot Q\right)\Big) 
+P_\mu \Big(\delta_{\alpha\gamma} \left(K\cdot Q-m^2\right)-K_\gamma Q_\alpha+K_\alpha Q_\gamma\Big)\\
&\hphantom{F\cdot\Big[}+P_\gamma K_\alpha Q_\mu
+P_\alpha K_\gamma Q_\mu +m^2 Q_\mu \delta_{\alpha\gamma}-Q_\mu \delta_{\alpha\gamma} P\cdot K+m^2 P_\alpha \delta_{\mu\gamma}
 +m^2 P_\gamma \delta_{\mu\alpha}\\
 &\hphantom{G\cdot\Big[}-m^2 K_\alpha \delta_{\mu\gamma}+m^2 K_\gamma \delta_{\mu\alpha}
 -m^2 Q_\alpha \delta_{\mu\gamma} +m^2 Q_\gamma \delta_{\mu\alpha}
 +Q_\alpha \delta_{\mu\gamma} P\cdot K -Q_\gamma \delta_{\mu\alpha} P\cdot K\\
 &\hphantom{G\cdot\Big[}-K_\alpha \delta_{\mu\gamma} P\cdot Q +K_\gamma \delta_{\mu\alpha} P\cdot Q
 -P_\alpha \delta_{\mu\gamma} K\cdot Q -P_\gamma \delta_{\mu \alpha} K\cdot Q \Big] \Big\}\, ,
\end{split}
\end{equation*}
where ${\cal S}$ stands for a full symmetrization of the subscript indices (without factorials).
This expression is very similar to that for the boson field obtained in the previous subsection,
with the proviso that now the Matsubara frequencies to be summed involve odd integers. We can 
then trace the bosonic procedure, first setting $Z$ to zero and defining the auxiliary function $S^F$:
\begin{equation*}
\begin{split}
C(X,Y,0)=C(\tau_1,\tau_2,{\bm x},{\bm y})=-\int \frac{\di^3 p }{(2\pi)^3}\int 
\frac{\di^3 q}{(2\pi)^3}\int \frac{\di^3 k }{(2\pi)^3}\e^{-\ii (\bm p+\bm k)\cdot x}
\e^{-\ii (\bm q -\bm p)}S^F(\bm p,\bm q, \bm k,\tau_1,\tau_2)
\end{split}
\end{equation*}
with:
\begin{equation*}
\begin{split}
S^F({\bm p},{\bm q},{\bm k},\tau_1, \tau_2)=\frac{1}{|\beta|^3}\sum_{\{p_n,q_n,k_n\}}&
\frac{\e^{\ii (p_n+\ii\mu)(\tau_1-\tau_2)+\ii (q_n+\ii\mu)\tau_2+\ii (k_n-\ii\mu)\tau_1}G_1(P^+,Q^+,K^-)}{({P^+}^2+m^2)({Q^+}^2+m^2)({K^-}^2+m^2)}\\
&+\frac{\e^{\ii (p_n-\ii\mu)\tau(\tau_1-\tau_2)+\ii (q_n-\ii\mu)\tau_2+\ii (k_n+\ii\mu)\tau_1}
G_2(P^-,Q^-,K^+)}{({P^-}^2+m^2)({Q^-}^2+m^2)({K^+}^2+m^2)}.
\end{split}
\end{equation*}
The sums over fermionic frequencies can be done by using a formula corresponding to
(\ref{sommabosoni}) for the boson field:
\begin{equation}
\label{sommafermioni}
\begin{split}
\frac{1}{\beta}\sum_{\{\omega_n\}}\frac{(\omega_n\pm \ii \mu)^k \e^{\ii (\omega_n\pm \ii \mu) \tau}}{(\omega_n\pm \ii \mu)^2+E^2}&=\frac{1}{2E}\left[
(-\ii E)^k\e^{\tau E}(\theta(-\tau)-n_F(E\pm\mu))+(\ii E)^k\e^{-E\tau}(\theta(\tau)-n_F(E\mp\mu))
\right]
\\
&=\frac{1}{2E}\sum_{s=\pm 1} (-\ii s E)^k\e^{\tau  s E}[\theta(- s \tau)-n_{F}(E\pm s\mu)]
\end{split}
\end{equation}
$n_{F}$ being the Fermi-Dirac distribution function:
\begin{equation*}
\begin{split}
n_{F}(E)=\frac{1}{\e^{|\beta|E}+1}.
\end{split}
\end{equation*}
Like for the boson case, $S^F$ comprises 8 terms:
\begin{equation}\label{sf1}
\begin{split}
&S^F({\bm p},{\bm q},{\bm k},\tau_1, \tau_2)=\frac{1}{8E_p E_q E_k }\sum_{s_1,s_2,s_3=\pm 1}
\e^{\tau_1 ( s_1 E_p +s_3 E_k )+\tau_2(s_2 E_q -s1E_q  )}\left\{ G_1\left(\tilde P(s_1) ,\tilde Q(s_2) ,\tilde K(s_3) \right)\right. \\
&
\left.
[\theta(- s_1( \tau_1-\tau_2))-n_{F}(E_p +s_1\mu)]  [\theta(- s_2 )-n_{F}(E_q +s_2\mu)]   
[\theta(- s_3 )-n_{F}(E_k -s_3\mu)]
+G_2\left(\tilde P(s_1) ,\tilde Q(s_2) ,\tilde K(s_3) \right)
\right.\\
&
\left. 
 [\theta(- s_1( \tau_1-\tau_2))-n_{F}(E_p -s_1\mu)]  [\theta(- s_2 )-n_{F}(E_q -s_2\mu)]   [\theta(- s_3 )-n_{F}(E_k +s_3\mu)]
\right\}
\end{split}
\end{equation}
with $\tilde P(s_1),\,\tilde Q(s_2)$ and $\tilde K (s_3)$ defined in (\ref{newmom}),
and the polynomial $G$ no longer depends on chemical potential after frequency
summation. Furthermore, 
$$
G_1\left(\tilde P(s_1) ,\tilde Q(s_2) ,\tilde K(s_3) \right)=
G_2\left(\tilde P(s_1) ,\tilde Q(s_2) ,\tilde K(s_3) \right)
$$ 
and so we can rewrite $S^F$ in eq.~(\ref{sf1}):
\begin{equation}\label{sf2}
\begin{split}
&S^F({\bm p},{\bm q},{\bm k},\tau_1, \tau_2)=\frac{1}{8E_p E_q E_k }
\sum_{s_1,s_2,s_3=\pm 1}\e^{\tau_1 ( s_1 E_p +s_3 E_k )+\tau_2(s_2 E_q -s_1E_q  )}
G_1\left(\tilde P(s_1) ,\tilde Q(s_2) ,\tilde K(s_3) \right)\\
&
\Big\{
[\theta(- s_1( \tau_1-\tau_2))-n_{F}(E_p +s_1\mu)]  [\theta(- s_2 )-n_{F}(E_q +s_2\mu)]   
[\theta(- s_3 )-n_{F}(E_k -s_3\mu)]+(\mu \to -\mu)
\Big\}.
\end{split}
\end{equation}
The spatial integrations are straightforward, and so we can finally write down the general
three-point function for the free Dirac field:
\begin{equation*}
\begin{split}
C^{\alpha\beta|\gamma\rho|\mu\nu|ij}=\frac{1}{|\beta|^2} \int_X \int_Y \media{{\rm T}_\tau
\h T^{\alpha\beta}(X)\h T^{\gamma\rho}(Y) \h T^{\mu\nu}(0)
}_{T,c} x^i y^j=\frac{-1}{|\beta|^2}\int^{|\beta|}_0\di \tau_1 \di\tau_2\int \frac{\di^3 p}{(2\pi)^3} \frac{\partial^2}{\partial k_i\partial q_j}S^F({\bm p},{\bm q},{\bm k},\tau_1, \tau_2)\Big |_{\substack{{\bm q}= {\bm p}\\ {\bm k}= -{\bm p} }}.
\end{split}
\end{equation*}
with $S^F$ given by the (\ref{sf2}).
The coefficients of the symmetrized stress-energy tensor can now be expressed by using the
relations (\ref{coefficientiC}) like in the boson case, as integrals of the second derivative 
of the Fermi-Dirac distribution function weighted with polynomials of momentum and mass:
\begin{equation}\label{fermcoeff}
\begin{split}
U_w&=\frac{1}{8\pi^2\beta^2}\int_0^\infty \frac{\di p}{E_p}  \left(n_F''(E_p-\mu)+n_F''(E_p+\mu)\right) p^2(p^2+m^2),\\
U_\alpha&=\frac{1}{24 \pi^2\beta^2}\int_0^\infty \frac{\di p}{E_p}  \left(n_F''(E_p-\mu)+n_F''(E_p+\mu)\right) 
 (p^2+m^2)^2 ,\\
W&=0,\\
A&=0,\\
G&=
\frac{1}{72\pi^2\beta^2}\int_0^\infty \frac{\di p}{E_p}  \left(n_F''(E_p-\mu)+n_F''(E_p+\mu)\right)  \left(4p^4+3m^2p^2\right),\\
D_\alpha&=
\frac{1}{72\pi^2\beta^2}\int_0^\infty \frac{\di p}{E_p}  \left(n_F''(E_p-\mu)+n_F''(E_p+\mu)\right)
 \left(p^4+3m^2p^2\right),\\
D_w&=
\frac{1}{24\pi^2\beta^2}\int_0^\infty \frac{\di p}{E_p}  \left(n_F''(E_p-\mu)+n_F''(E_p+\mu)\right)p^4,
\end{split}
\end{equation}
where $n_F''(E_p\pm\mu)=\di^2 n_F(E_p\pm\mu)/\di E_p^2$.

Also in the fermionic case, it can be shown that the coefficients (\ref{fermcoeff}) 
fulfill the relations (\ref{coeffrel}).

\section{The vector current}

Also conserved currents can receive corrections in general thermodynamic equilibrium 
with non-vanishing thermal vorticity (\ref{densop2}). For a vector current $\h j^\mu(x)$, 
the expansion (\ref{tauexp3}) yields: 
\begin{equation}\label{current}
\begin{split}
j^\mu(x) &=\media {\h j^\mu(x)}_{\beta(x)}-\frac{\alpha_\rho}{|\beta|}\int_0^{|\beta|}
\di \tau \media{{\rm T}_\tau \left(\h K^{\rho}_{-\ii \tau u}\,\h j^\mu(0)\right)}_{\beta(x),c}
-\frac{w_\rho}{|\beta|}\int_0^{|\beta|}\di \tau \media{{\rm T}_\tau\left(\h J^{\rho}_{-\ii \tau u}\,
\h j^\mu(0)\right)}_{\beta(x),c}\\
&+\frac{\alpha_\rho\alpha_\sigma}{2|\beta|^2}\int_0^{|\beta|}\di \tau_1\di \tau_2
\media{{\rm T}_\tau\left(\h K^{\rho}_{-\ii \tau_1 u }\,\h K^{\sigma}_{-\ii \tau_2 u}\,\h j^\mu(0)\right)
}_{\beta(x),c}+\frac{w_\rho w_\sigma}{2|\beta|^2}\int_0^{|\beta|}\di \tau_1\di \tau_2
\media{{\rm T}_\tau\left(\h J^{\rho}_{-\ii \tau_1 u }\,\h J^{\sigma}_{-\ii \tau_2 u}\,\h j^\mu(0)\right)}_{\beta(x),c}
\\
&+\frac{\alpha_\rho w_\sigma}{2|\beta|^2}\int_0^{|\beta|}\di \tau_1\di \tau_2
\media{{\rm T}_\tau\left(\{\h K^{\rho}_{-\ii \tau_1 u }\,,\h J^{\sigma}_{-\ii \tau_2 u}\}\,\h j^\mu(0)\right)}_{\beta(x),c}
+\mathcal{O}(\varpi^3).
\end{split}
\end{equation}
The leading order term $\media {\h j^\mu(x)}_{\beta(x)}$ is simply the homogeneous equilibrium
current $n u^\mu$ where $n$ is the proper charge density. The first order corrections in $\alpha$ 
and $w$ are zero due to the time-reversal and parity symmetries, just like for the stress-energy tensor;
hence, again, the first non-vanishing corrections are quadratic in thermal vorticity. 
The invariance under rotation selects only three allowed tensor combinations with $\alpha$ and $w$:
\begin{equation}\label{currdec}
 j_\mu(x) = n\, u_\mu  -(\alpha^2 N_\alpha+ w^2 N_w )\,u_\mu + G_V\gamma_\mu + {\cal O}(\varpi^3)
\end{equation}
where $n$ is mean value of the charge density at the homogeneous equilibrium. By comparing
the eqs.~(\ref{currdec}) and (\ref{current}) and taking into account the rotational invariance, 
we can identify in the local rest frame the following formulae for $N_\alpha$, $N_w$ and $G_V$:
\begin{equation}\label{currcoeff}
\begin{split}
 N_\alpha &=\frac{1}{2|\beta|^2}\int_0^{|\beta|}\di \tau_1\di \tau_2
\media{{\rm T}_\tau\left(\h K^3_{-\ii \tau_1 }\,\h K^3_{-\ii \tau_2 }\,\h j^0(0)\right)}_{T,c}\, ,\\
 N_w&=\frac{1}{2|\beta|^2}\int_0^{|\beta|}\di \tau_1\di \tau_2
\media{{\rm T}_\tau\left(\h J^3_{-\ii \tau_1 }\,\h J^3_{-\ii \tau_2 }\,\h j^0(0)\right)}_{T,c}\, ,\\
 G_V&=\frac{1}{2|\beta|^2}\int_0^{|\beta|}\di \tau_1\di \tau_2
\media{{\rm T}_\tau \left(\{\h K^1_{-\ii \tau_1 },\h J^2_{-\ii \tau_2 }\}\,\h j^3(0)\right)}_{T,c}\, .
\end{split}
\end{equation}
The right hand sides of the above equalities involve the three-point thermal functions of 
the stress-energy tensor (twice) and the current operator. Defining:
\begin{equation}\label{bblockcurrent}
C^{\alpha\beta|\gamma\rho|\mu|ij}=\frac{1}{|\beta|^2}\int_0^{|\beta|}\di \tau_1\di \tau_2\int\di^3x\di^3y \media{{\rm T}_\tau \left(\h T^{\alpha\beta}( \tau_1 ,{\bm x})\h T^{\gamma\rho}( \tau_2 ,{\bm y}) \h j^{\mu}(0)\right)}_{T,c} x^i y^j
\end{equation}
the coefficients in eq.~(\ref{currcoeff}) can be obtained as linear combinations of (\ref{bblockcurrent}):
\begin{equation}\label{currcoeffC}
\begin{split}
 N_\alpha &=\frac{1}{2}C^{00|00|0|33}\, ,\\
 N_w&=\frac{1}{2}(C^{01|01|0|22}-C^{01|02|0|21}-C^{02|01|0|12}+C^{02|02|0|11})\, ,\\
 G^V&=-\frac{1}{2}(C^{00|03|3|11}-C^{00|01|3|13}+C^{03|00|3|11}-C^{01|00|3|31}).
\end{split}
\end{equation}
%

\subsection{Free complex scalar field}

The current of the free scalar field reads:
\begin{equation*}
\h j_\mu=\ii (\wpsi^\dagger \partial_{\mu}\wpsi-\wpsi\partial_{\mu}\wpsi^\dagger)
\end{equation*}
and its Euclidean counterpart:
\begin{equation*}
   \h j_\mu=\ii (-\ii)^{\delta_{0\mu}} \left( \wpsi^\dagger\partial_\mu \wpsi-\partial_\mu 
   \wpsi^\dagger\wpsi \right).
\end{equation*}
The general coefficient in eq.~(\ref{bblockcurrent}) can be calculated by using the point 
splitting procedure:
\begin{equation}\label{3pfcurr}
\begin{split}
C_{\text{curr.}}(X,Y,Z)=\media{{\rm T}_\tau \h T^{\alpha\beta}(X)\h T^{\gamma\rho}(Y) \h j^{\mu}(Z)}_{T,c}
=&\lim_{\substack{X_1,X_2\to X \\ Y_1,Y_2\to Y\\ 
Z_1,Z_2\to Z}}\mathcal{D}^{\alpha\beta}(\partial_{X_1},\partial_{X_2}) \mathcal{D}^{\gamma\rho}(\partial_{Y_1},\partial_{Y_2}) \mathcal{J}^{\mu}(\partial_{Z_1},\partial_{Z_2})
\\&\times \media{{\rm T}_\tau \wpsi^\dagger(X_1)\wpsi(X_2)\wpsi^\dagger(Y_1)\wpsi(Y_2)\wpsi^\dagger(Z_1)\wpsi(Z_2)}_{T,c},
\end{split}
\end{equation}
where the differential operator $\mathcal{D}^{\alpha \beta}$ is defined in (\ref{polinomio}). The
operator $\mathcal{J}^\mu$ is found inserting  the current operator into the correlation function:
\begin{equation*}
\mathcal{J}^\mu(\partial_X,\partial_Y)=\ii (-\ii)^{\delta_{0\mu}} \left( \partial^\mu_Y -\partial^\mu_X \right).
\end{equation*}
Thus:
\begin{equation}\label{3pfcurrf}
\begin{split}
C_{\text{curr.}}^{\alpha\beta|\gamma\rho|\mu|ij}=\frac{1}{|\beta|^2} \int_X \int_Y \media{ {\rm T}_\tau
\h T^{\alpha\beta}(X)T^{\gamma\rho}(Y) \h j^{\mu}(0)
}_{T,c} x^i y^j=\frac{-1}{|\beta|^2}\int^{|\beta|}_0\di \tau_1 \di\tau_2\int \frac{\di^3 p}{(2\pi)^3} 
\frac{\partial^2}{\partial k_i\partial q_j}S({\bm p},{\bm q},{\bm k},\tau_1, \tau_2)
\Big|_{\substack{{\bm q}= {\bm p}\\ {\bm k}= -{\bm p} }},
\end{split}
\end{equation}
where the function $S$ is now:
\begin{equation}\label{scurrf}
\begin{split}
S({\bm p},{\bm q},{\bm k},\tau_1, \tau_2)=\frac{1}{|\beta|^3}\sum_{p_n,q_n,k_n}\frac{\e^{\ii 
P^+(\tau_1-\tau_2)+\ii Q^+\tau_2+\ii K^-\tau_1}H_1(P^+,Q^+,K^-)}{(P^{+2}+m^2)(Q^{+2}+m^2)
(K^{-2}+m^2)}+\frac{\e^{\ii P^-(\tau_1-\tau_2)+\ii Q^-\tau_2+\ii K^+\tau_1}H_2(P^-,Q^-,K^+)}
{(P^{-2}+m^2)(Q^{-2}+m^2)(K^{+2}+m^2)}
\end{split}
\end{equation}
and the polynomials in the momenta are:
\begin{equation*}
\begin{split}
H_1(P^+,Q^+,K^-)=& \mathcal{D}^{\alpha\beta}(\ii P^+,\ii K^-) \mathcal{D}^{\gamma\rho}(\ii Q^{+},-\ii P^{+}) 
\mathcal{J}^{\mu}(-\ii K^{-},-\ii Q^+),\\
H_2(P^-,Q^-,K^+)=&\mathcal{D}^{\alpha\beta}(\ii K^+,\ii P^-) \mathcal{D}^{\gamma\rho}(-\ii P^{-},\ii Q^{-}) 
\mathcal{J}^{\mu}(-\ii Q^{-},-\ii K^+) .
\end{split}
\end{equation*}
Since the polynomial $\mathcal{J}^\mu(X,Y)$ is antisymmetric by argument swap, after 
the frequency summation the two polynomials $H_1$ and $H_2$ no longer depend on the 
chemical potential and are opposite, i.e. $H_1=-H_2$.

Then, summing over Matsubara frequencies and reminding definition (\ref{newmom}):
\begin{equation*}
\begin{split}
&S({\bm p},{\bm q},{\bm k},\tau_1, \tau_2)=\frac{1}{8E_p E_q E_k }
\sum_{s_1,s_2,s_3=\pm 1}\e^{\tau_1 ( s_1 E_p +s_3 E_k )+\tau_2(s_2 E_q -s1 E_p )}
\left\{ H_1\left(\tilde P(s_1) ,\tilde Q(s_2) ,\tilde K(s_3) \right)\right. \\
&
\left.
[\theta(- s_1( \tau_1-\tau_2))+n_{B}(E_p +s_1\mu)]  [\theta(- s_2 )+n_{B}(E_q +s_2\mu)]   
[\theta(- s_3 )+n_{B}(E_k -s_3\mu)]+H_2\left(\tilde P(s_1) ,\tilde Q(s_2) ,\tilde K(s_3) \right)
\right.\\
&
\left. 
 [\theta(- s_1( \tau_1-\tau_2))+n_{B}(E_p -s_1\mu)]  [\theta(- s_2 )+n_{B}(E_q -s_2\mu)]   
 [\theta(- s_3 )+n_{B}(E_k +s_3\mu)]
\right\}\\
&=\frac{1}{8E_p E_q E_k }\sum_{s_1,s_2,s_3=\pm 1}\e^{\tau_1 ( s_1 E_p +s_3 E_k )+
\tau_2(s_2 E_q -s_1 E_p)}H_1\left(\tilde P(s_1) ,\tilde Q(s_2) ,\tilde K(s_3) \right) \\
&
\{
[\theta(- s_1( \tau_1-\tau_2))+n_{B}(E_p +s_1\mu)]  [\theta(- s_2 )+n_{B}(E_q +s_2\mu)]   
[\theta(- s_3 )+n_{B}(E_k -s_3\mu)]
-(\mu\to -\mu)\}.
\end{split}
\end{equation*}

Taking the derivative of $S$ respect to $\bm q$ and $\bm k$, integrating over $\tau_1$ and $\tau_2$, 
according to eq.~(\ref{3pfcurrf}), the coefficients (\ref{currcoeffC}) turn out to be:
\begin{equation}
\begin{split}\label{boscurrcoeff}
 N_\alpha =& \frac{m^2}{48\pi^2\beta^2}\int_0^\infty \di p \left(n_B''(E_p-\mu)-n_B''(E_p+\mu)\right),\\
 N_w =& \, 0 ,\\
 G^V =&-\frac{1}{12\pi^2\beta^2}\int_0^\infty \di p \left(n_B''(E_p-\mu)-n_B''(E_p+\mu)\right)p^2. 
\end{split}
\end{equation}
%

\subsection{Free Dirac field } 

The vector current of Dirac field $\h{j}^\mu = \wbPsi \gamma^\mu\wPsi $ in its Euclidean 
version reads:
\begin{equation*}
\h j^\mu=(-\ii)^{1-\delta_{0\mu}}\wbPsi \tilde\gamma^\mu\wPsi ,
\end{equation*}
where $\tilde \gamma$ are the Euclidean gamma matrices. The generic coefficient (\ref{bblockcurrent}) 
in this case can be written as:
\begin{equation}\label{dir3pf}
\begin{split}
C^{\text{curr.}}(X,Y,Z)&=\media{{\rm T}_\tau \h T^{\alpha\beta}(X)\h T^{\gamma\rho}(Y) \h j_V^{\mu}(Z)
}_{T,c} =\\
\lim_{\substack{X_1,X_2\to X \\ Y_1,Y_2\to Y\\ 
Z_1,Z_2\to Z}}&\mathcal{D}^{\alpha\beta}(\partial_{X_1},\partial_{X_2})_{ab} \mathcal{D}^{\gamma\rho}(\partial_{Y_1},\partial_{Y_2})_{cd} \mathcal{J}^{\mu}(\partial_{Z_1},\partial_{Z_2})_{ef}
 \media{{\rm T}_\tau \wbPsi_a(X_1)\wPsi_b(X_2)\wbPsi_c(Y_1)\wPsi_d(Y_2)\wbPsi_e(Z_1)\wPsi_f(Z_2)}_{T,c}
\end{split}
\end{equation}
where the matrix associated to the stress-eenrgy tensor $\mathcal{D}^{\alpha\beta}(\partial_{X_1},
\partial_{X_2})_{ab}$ is defined in eq.~(\ref{fermD}), and $\mathcal{J}^{\mu}(\partial_{Z_1},
\partial_{Z_2})_{ef}$ corresponding to the vector current is:
\begin{equation*}
  \mathcal{J}^{\mu}(\partial_{Z_1},\partial_{Z_2})_{ef}= (-\ii)^{1-\delta_{0\mu}} (\tilde\gamma^\mu)_{ef}\, .
\end{equation*}
The function $S$ analogous to that in eq.~(\ref{scurrf}) is:
\begin{equation*}
\begin{split}
S^F({\bm p},{\bm q},{\bm k},\tau_1, \tau_2)=-\frac{1}{|\beta|^3}\sum_{\{p_n,q_n,k_n\}}
\frac{\e^{\ii P^+(\tau_1-\tau_2)+\ii Q^+\tau_2+\ii K^-\tau_1}M_1(P^+,Q^+,K^-)}{(P^{+2}+m^2)(Q^{+2}+m^2)
(K^{-2}+m^2)}+\frac{\e^{\ii P^-\tau(\tau_1-\tau_2)+\ii Q^-\tau_2+\ii K^+\tau_1}M_2(P^-,Q^-,K^+)}
{(P^{-2}+m^2)(Q^{-2}+m^2)(K^{+2}+m^2)}
\end{split}
\end{equation*}
where, in this case, the vertex functions corresponding to the stress-energy tensor and the
vector current insertion into the correlation function (\ref{dir3pf}) are denoted as $M_1$ 
and $M_2$:
\begin{equation*}
\begin{split}
M_1(P^+,Q^+,K^-) &=\tr\left[\mathcal{D}_{\alpha\beta}(\ii K^-,\ii P^+)(-\ii\slashed P^++m)
\mathcal{D}_{\gamma\rho}(-\ii P^+,\ii Q^+)(-\ii\slashed Q^++m)
\mathcal{J}_{\mu}(-\ii Q^+,-\ii K^-)(\ii\slashed K^-+m)\right],
\\
M_2(P^-,Q^-,K^+)& =\tr\left[
\mathcal{D}_{\alpha\beta}(\ii P^-,\ii K^+)(-\ii\slashed K^++m)\mathcal{J}_{\mu}(-\ii K^+,-\ii Q^-)(\ii\slashed Q^-+m)\mathcal{D}_{\gamma\rho}(\ii Q^-,-\ii P^-)(\ii\slashed P^++m)
\right].
\end{split}
\end{equation*}
The sum over frequencies yields:
\begin{equation*}
\begin{split}
&S^F({\bm p},{\bm q},{\bm k},\tau_1, \tau_2)=-\frac{1}{8E_p E_q E_k }\sum_{s_1,s_2,s_3=\pm 1}\e^{\tau_1 ( s_1 E_p +s_3 E_k )+\tau_2(s_2 E_q -s_1E_p  )}M_1\left(\tilde P(s_1) ,\tilde Q(s_2) ,\tilde K(s_3) \right)\\
&
\Big\{
[\theta(- s_1( \tau_1-\tau_2))-n_{F}(E_p +s_1\mu)]  [\theta(- s_2 )-n_{F}(E_q +s_2\mu)]   [\theta(- s_3 )-n_{F}(E_k -s_3\mu)]-(\mu \to -\mu)
\Big\}
\end{split}
\end{equation*}
where, like in the boson case, the two functions $M_1$ and $M_2$ turn out to be opposite:
\begin{equation*}
\begin{split}
 &M_1(P,Q,K)=-M_2(P,Q,K)={\cal S}_{\alpha\beta\gamma\rho} \Bigr\{
 \ii (K^\beta-P^\beta) (P^\rho+Q^\rho) \Big[K^\mu \big(P^\alpha Q^\gamma + P^\gamma 
 Q^\alpha - \delta^{\gamma\alpha} (m^2 + P\cdot Q)\big)\\
 &+P^\mu \left(\delta^{\gamma\alpha} \left(K\cdot Q-m^2\right)+K^\alpha Q^\gamma-K^\gamma 
 Q^\alpha\right)+Q^\mu \big(\delta^{\gamma\alpha}(m^2- P\cdot K)+K^\gamma P^\alpha+ K^\alpha P^\gamma\big)\\
 &+(K^\gamma \delta^{\mu\alpha} - K^\alpha \delta^{\mu\gamma})(m^2 + P\cdot Q)+
 (P^\gamma \delta^{\mu\alpha} + P^\alpha \delta^{\mu\gamma})(m^2 - K\cdot Q)
 +(Q^\gamma \delta^{\mu\alpha} - Q^\alpha \delta^{\mu\gamma})(m^2 - K\cdot P)\Big]\Bigl\}.
\end{split}
\end{equation*}
After performing the integrations in $\tau$, taking the derivative with respect to the 
momenta $\bm q $ and $\bm k$ and setting the appropriate indices according the 
eq.~(\ref{currcoeffC}), the coefficients for the Dirac field are finally obtained:
\begin{equation}
\begin{split}\label{fermcurrcoeff}
 N_\alpha =&\frac{1}{24 \pi^2 \beta^2}\int_0^\infty \di p \left(n_F''(E_p -\mu)-
 n_F''(E_p +\mu)\right)(3p^2+m^2) ,\\
 N_w =&- \frac{1}{8\pi^2\beta^2}\int_0^\infty \frac{\di p}{E_p } 
 \left(n_F'(E_p -\mu)-n_F'(E_p +\mu)\right)(2 p^2+ m^2),\\
 G^V =&-\frac{1}{12\pi^2\beta^2}\int_0^\infty \frac{\di p}{E_p }
 \left(n_F'(E_p -\mu)-n_F'(E_p +\mu)\right)(2p^2+m^2). 
\end{split}
\end{equation}
%

\section{The axial current}

It is also worth studying the lowest-order expression of the mean axial current $\media{\h j_A^\mu}$ at thermodynamic equilibrium with vorticity and acceleration. For the free field without interactions, 
the axial current $\h j_A^\mu=\wbPsi  \gamma^\mu\gamma^5\wPsi$ is conserved only in the massless 
limit, being, as it is well known:
\begin{equation*}
\begin{split}
\partial_\mu \h j_A^\mu=2m \ii \wbPsi \gamma^5 \wPsi.
\end{split}
\end{equation*}
Unlike the vector current, because of its properties under space reflection and time-reversal,
its mean value vanishes at homogeneous thermodynamic equilibrium, but it has a first-order
correction proportional to vorticity:
\begin{equation}\label{axcurr}
j_A^\mu(x) = \media{\h j^\mu_A(x)}=-\frac{w_\rho}{|\beta|}\int_0^{|\beta|}\di \tau \media{
{\rm T}_\tau\left(\h J^{\rho}_{-\ii \tau u}\,\h j^\mu_A(0)\right)}_{\beta(x),c} + 
\mathcal{O}(\varpi^2) = w^\mu W^A + \mathcal{O}(\varpi^2). 
\end{equation}

The coefficient $W^A$ can be calculated from the two-point function between the angular momentum 
operator and the axial current operator; in the rest frame its formula is: 
\begin{equation}
\label{coeffw}
\begin{split}
W^A&=\frac{1}{|\beta|}\int_0^{|\beta|}\di \tau \media{{\rm T}_\tau\left(\h J^{3}_{-\ii \tau}
\h j^{3,A}(0)\right)}_{T,c} \\
&=\frac{1}{|\beta|}\int_0^{|\beta|}\di \tau\int \di^3 x \left(\media{{\rm T}_\tau\h T^{02}(\tau ,\bm x)
\h j^3_{A}(0)}_{T,c}x^1- \media{{\rm T}_\tau\h T^{01}(\tau ,\bm x)\h j^3_{A}(0)}_{T,c}x^2\right)
\\
&\equiv C_{\text{axial}}^{02|3|1}-C_{\text{axial}}^{01|3|2}.
\end{split}
\end{equation}
To calculate the latter two terms, as usual, we write the correlation functions in the Euclidean 
form. The Euclidean axial current reads:
\begin{equation*}
\h j_A^\mu(X)=(-\ii)^{1-\delta_{0\mu}}\wbPsi(X) \tilde\gamma^\mu \tilde\gamma^5\wPsi(X) =
\lim_{X_1,X_2\to X}(-\ii)^{1-\delta_{0\mu}} (\tilde\gamma^\mu \tilde\gamma^5)_{ab} 
\wbPsi_a(X_1)\wPsi_b(X_2)=\mathcal{J}_A^{\mu}(\partial_{X_1},\partial_{X_2})_{ab}\wbPsi_a(X_1)\wPsi_b(X_2)
\end{equation*}
where $\tilde \gamma^5$ is defined as:
\begin{equation*}
\begin{split}
\tilde \gamma^5=\tilde \gamma^0\tilde \gamma^1\tilde \gamma^2\tilde \gamma^3
\end{split}
\end{equation*}
and it is equal to the usual $\gamma^5$ matrix. Then, we can write:
\begin{equation*}
\begin{split}
\media{{\rm T}_\tau\h T^{\mu\nu}(X)\h j^\alpha_{A}(Y)}_{T,c}&=\lim_{\substack{X_1,X_2\to X \\ Y_1,Y_2\to Y}}
\mathcal{D}^{\mu\nu}(\partial_{X_1},\partial_{X_2})_{ab} \mathcal{J}_A^{\alpha}(\partial_{Y_1},\partial_{Y_2})_{cd}
 \media{{\rm T}_\tau\wbPsi_a(X_1)\wPsi_b(X_2)\wbPsi_c(Y_1)\wPsi_d(Y_2)}_{T,c}.
\end{split}
\end{equation*}
By using the Wick theorem and the momentum representation of the Dirac propagator we can write:
\begin{equation*}
\begin{split}
\media{{\rm T}_\tau\h T^{\mu\nu}(X)\h j^\alpha_{A}(0)}_{T,c}&=-\SumInt_{\{P,Q\}}A(\ii Q^-,\ii P^+)\Delta(P^+)\Delta(Q^-)\e^{\ii (P+Q)\cdot X},
\end{split}
\end{equation*}
where the function $A$ results from of the composite operator in the correlation functions, 
in this case:
\begin{equation}\label{Afunz}
\begin{split}
A(\ii P, \ii Q)&=\tr\left[\mathcal{D}^{\mu\nu}(\ii Q,\ii P )(-\ii \slashed P+m)
\mathcal{J}_A^{\alpha}(-\ii P,-\ii Q)(\ii \slashed Q+m) \right]\\
&=(\ii P^\mu-\ii Q^\mu)\epsilon^{\alpha\nu\rho \sigma}(\ii P_\rho)(\ii Q_\sigma)+(\ii P^\nu-\ii Q^\nu)
\epsilon^{\alpha\mu\rho \sigma}(\ii P_\rho)(\ii Q_\sigma).
\end{split}
\end{equation}
After having defined $S_{\textrm{axial}}$  in the same fashion as in previous sections:
\begin{equation*}
\begin{split}
\media{{\rm T}_\tau
\h T^{\mu\nu}(X)\h j^\alpha_{A}(0)}_{T,c}=-\int\frac{\di^3 p}{(2\pi)^3}\int\frac{\di^3 q}{(2\pi)^3}
\e^{-\ii (\bm p+\bm q)\cdot \bm x} S_{\textrm{axial}}(\bm p,\bm q,\tau ),
\end{split}
\end{equation*}
we get:
\begin{equation*}
\begin{split}
S_{\mathrm{axial}}(\bm p, \bm q ,\tau)=\frac{1}{\beta^2}\sum_{\{p_n,q_n\}}\frac{\e^{\ii (p_n+\ii \mu)\tau}}{(p_n+\ii \mu)^2+E^2_p}\frac{\e^{\ii (q_n-\ii \mu)\tau}}{(q_n-\ii \mu)^2+E^2_p}A(\ii (p_n+\ii\mu),\ii \bm  p,\ii (q_n-\ii \mu),\ii \bm q).
\end{split}
\end{equation*}
In the above expression, the two sums over frequencies are independent; using (\ref{sommafermioni}):
\begin{equation*}
\begin{split}
 S_{\mathrm{axial}}(\bm p, \bm q ,\tau)=\frac{1}{4E_pE_q} \Big\{
 &A(-E_p,\ii \bm p,-E_q,\ii \bm q )\e^{-(E_p+E_q)\tau}[1-n_F(E_p-\mu)][1-n_F(E_q+\mu)]
 \\-& A(-E_p,\ii \bm p,E_q,\ii \bm q )\e^{(-E_p+E_q)\tau}[1-n_F(E_p-\mu)]n_F(E_q+\mu)
\\ - &A(E_p,\ii \bm p,-E_q,\ii \bm q )\e^{(E_p-E_q)\tau}n_F(E_p-\mu)[1-n_F(E_q+\mu)]
\\+ &A(E_p,\ii \bm p,E_q,\ii \bm q )\e^{(E_p+E_q)\tau}n_F(E_p-\mu)n_F(E_q+\mu)
 \Big\}
 \end{split}
\end{equation*}
with the function $A$ defined in (\ref{Afunz}). Recalling the definition of $W^A$ in the eq.~(\ref{coeffw}) 
the integration over the spatial coordinates can be done as a derivative respect to a loop momenta:
\begin{equation*}
\begin{split}
C_{\mathrm{axial}}^{\mu\nu|\alpha | i}=\frac{1}{|\beta|}\int_0^{|\beta|}\di \tau\int \di^3 x 
\media{{\rm T}_\tau\h T^{\mu\nu}(X)\h j^\alpha_{A}(0)}_{T,c}x^i=-\ii \frac{1}{|\beta|}\int_0^{|\beta|}\di \tau \int \frac{\di^3 p}{(2\pi)^3}\frac{\partial}{\partial q^i} S_{\mathrm{axial}}(\bm p, \bm q ,\tau)\Big|_{\bm q=-\bm p}.
\end{split}
\end{equation*}
Choosing the suitable indices as in eq.~(\ref{coeffw}), taking the derivative with respect 
to the momentum $\bm q$ and integrating over both $\tau$ and the angles, we finally obtain 
the coefficient $W^A$ for the free Dirac field:
\begin{equation}\label{coeffw2}
W^A=\frac{1}{2 \pi^2 |\beta| }\int_0^\infty \frac{\di p}{E_p} \left(n_F(E_p-\mu)+n_F(E_p+\mu)\right)(2p^2+m^2).
\end{equation}

This result can be checked by using the conservation of the axial current in the massless case.
In general, the divergence of the mean axial current, at first order in $w$:
\begin{equation*}
\begin{split}
\partial_\mu j_A^\mu = \partial_\mu(w^\mu W^A) = 
- \alpha \cdot w \left(3\frac{W^A}{|\beta|} +\frac{\partial}{\partial |\beta|}W^A \right)
\end{split}
\end{equation*}
where we have used the fact that $W^A$ depends only on the magnitude of $\beta$ and the 
expressions of the gradients of $w^\mu$ that can be found in Appendix \ref{appb}. 
Specifically, the above combination gives rise to:
\begin{equation*}
\begin{split}
  \left(3\frac{W^A}{|\beta|} +\frac{\partial}{\partial \beta}W^A \right)=\frac{m^2}{2 \pi^2 \beta^2 }
  \int_0^\infty \frac{\di p}{E_p} \left(n^{\prime}_F(E_p-\mu)+n^\prime_F(E_p+\mu)\right),
\end{split}
\end{equation*}
which is manifestly vanishing for a massless Dirac field.

\subsection{Discussion: axial current and anomalies}\label{axdiscu}

The equation (\ref{axcurr}) states that in rotating fermion gas there is a non-vanishing axial 
current and consequently the right and left chiral fermions get separated. This current can 
be pictorially understood with a simple argument, which is strictly valid only for massless 
particles. In a rotating system the fermions spin tend to align with the direction of rotation 
$\bm \omega$ independently of their charge \cite{Becattini:2007nd}. The right handed particles 
have their momentum aligned with the spin, consequently will move in the direction of the 
spin, i.e. we get a net right handed particles flow in the direction of $\bm{\omega}$. On 
the other hand the left handed particles will move in the opposite direction, giving a net 
left handed particles flow opposite to $\bm{\omega}$. Together these flows give an axial 
current $\bm{j}\ped{A}=n_R \bm{v}_R-n_L \bm{v}_L\propto \bm{\omega}$. The very reason 
of the non-vanishing axial current is simply that it is allowed by the symmetry of the statistical 
operator, as $\bm{\h j}_A$ has the same properties of the angular momentum operator $\h {\bf J}$ 
under reflection, time reversal and charge conjugation; so:
$$
  {\bf j}_A = \frac{1}{Z} \tr ({\h {\bf j}_A} \exp[-\h H/T_0 + \omega \h J_z /T_0]) \ne 0.
$$
Using the method described in the next section~\ref{limcase} one can obtain the value of 
the coefficient $W\apic{A}$ (\ref{coeffw2}) for $m=0$:
\begin{equation}\label{AVEnomass}
  \bm{j}\ped{A}=\frac{W\apic{A}}{T}\,\bm{\omega}=
  \left( \frac{T^2}{6}+\frac{\mu^2}{2\pi^2}\right)\bm{\omega}.
\end{equation}

This result is precisely the same found in recent calculations of the mean axial current related
to the chiral vortical effect (CVE), which is the onset of a vector current along the vorticity 
due to the axial anomaly (see~\cite{Kharzeev:2015znc} and references therein). It should be
pointed out, though, that the onset of an axial current along vorticity is conceptually 
a distinct phenomenon, so we denote it by axial vortical effect (AVE) following ref.~\cite{Kalaydzhyan:2014bfa}.
The equality of the coefficients found with our method and the anomaly-related
method is a consequence of the equality of the Kubo formulae 
\cite{Amado:2011zx,Landsteiner:Holo,ChiralKinetic,Landsteiner:2011cp,Avkhadiev:2017fxj,Golkar:2012kb} obtained with either approach.
In view of the explicit absence of anomaly in our calculation - we deal with free fields in 
flat space-time from the outset and we ignore gauge interactions - one may wonder whether this 
equality is accidental or if our derivation is somehow equivalent to the anomalous one for some
reason.

The AVE was originally addressed for neutrinos emitted by rotating black hole in ref.~\cite{VilenkinCVE}.
Lately, as has been mentioned, this effect was addressed in the context of CVE and quantum 
anomalies \cite{PhysRevD.70.074018,Son:2009tf,Sadofyev:2010pr}. Several terms were found to 
contribute to the proportionality coefficient between ${\bf j}_A$ and ${\bf \omega}$, in the 
massless limit: a term proportional to the chiral potential $\mu_5^2$, also confirmed in holographic 
models \cite{TorabianYee,Erdmenger:2008rm,Banerjee:2008th,Amado:2011zx}; a term proportional to $T^2$ \cite{Neiman:2010zi,Landsteiner:Holo,ChiralKinetic} whose existence was attributed 
to the gravitational anomaly~\cite{Jensen:2012kj,Landsteiner:2011cp,Avkhadiev:2017fxj}
and to the modular anomaly~\cite{Golkar:2015oxw}.
The relation to anomalies is made stronger in ref.~\cite{Golkar:2012kb} where it is proven that the CVE does 
not receive corrections from a Yukawa type interactions.
Certainly, when dynamical degrees of freedom are considered instead of external fields, all 
kind of anomalous transport coefficients (CVE and AVE) are no more related nor constrained by
anomalies,
as first shown in~\cite{Golkar:2012kb} and then in~\cite{Jensen:2013vta,Braguta:2014gea}.
It is also worth pointing out that in fact, all calculations of the $T^2$ term in eq.~(\ref{AVEnomass}) in refs.~\cite{VilenkinCVE,Landsteiner:Holo,ChiralKinetic} give the same result, i.e. $T^2/6$ like 
in eq.~(\ref{AVEnomass}).
We would 
like to point out that in the derivation of Vilenkin \cite{VilenkinCVE} it is clear that the 
effect is caused by the modified density operator in the presence of rotation, exactly like in 
our case, and indeed we recover the same result in eq.~(\ref{AVEnomass}).
We also stress that 
in this framework, the coefficient $W_A$ is non-vanishing also for massive fields, as it turns 
out from the general formula (\ref{coeffw2}), when the chiral symmetry is explicitly broken.
Furthermore, it is likely that the addition of a term $\mu_5 \h Q_5/T$, where $\h Q_5$ is a 
conserved axial charge, in the exponent of the statistical operator (\ref{densop}), 
will lead to a term $\mu_5^2/2\pi^2$ as found in~\cite{PhysRevD.70.074018,Son:2009tf,TorabianYee,Erdmenger:2008rm,
Banerjee:2008th,Amado:2011zx,ChiralKinetic}.

Finally, it should also be pointed out that the formula (\ref{coeffw2}) was implicitly obtained 
in ref.~\cite{Becattini:2014ova} by means of the relativistic single-particle distribution
function of particles with spin $1/2$. Therein, the mean spin tensor of the free Dirac field:
\begin{equation*}
\spt^{\lambda,\rho\sigma}=\frac{\ii}{8}\media{\wbPsi\{\gamma^\lambda,[\gamma^\rho,\gamma^\sigma]\}\wPsi},
\end{equation*}
was found to be, with a non-vanishing thermal vorticity $\varpi$:
\begin{equation}\label{spintensor}
 \spt^{\lambda,\rho\sigma}=\iota\left( u^\lambda\varpi^{\rho\sigma}+u^\sigma\varpi^{\lambda\rho}+u^\rho\varpi^{\sigma\lambda}\right)
\end{equation}
where:
\begin{equation*}
 \iota\equiv -\frac{1}{T}\frac{\partial F}{\partial\beta^2};\qquad\qquad
 F(\beta^2,\zeta)\equiv \frac{1}{(2\pi)^3} 
 \int \frac{\di^3 p}{E_p}\left(\frac{1}{\E^{|\beta|E_p-\zeta}+1}+\frac{1}{\E^{|\beta|E_p+\zeta}+1}\right).
\end{equation*}
Noticing that the derivative of $\beta^2$ inside $F$ can be recast as a derivative over energy of 
the Fermi distribution $n_F$, after integration by parts it can be shown that 
$$
  2\iota = \frac{1}{2 \pi^2 |\beta| }\int_0^\infty \frac{\di p}{E_p} 
  \left(n_F(E_p-\mu)+n_F(E_p+\mu)\right)(2p^2+m^2)= W^A
$$
according to eq.~(\ref{coeffw2}). Since the axial current is proportional to the dual of the spin tensor $\mathcal{S}$:
\begin{equation}\label{spintenax}
\h j_\mu^A(x) = \wbPsi \gamma_\mu \gamma^5 \wPsi =-\frac{1}{3}\epsilon_{\mu\lambda\rho\sigma}
\wspt^{\lambda,\rho\sigma},
\end{equation}
using the eq.~(\ref{spintenax}) and (\ref{spintensor}) the mean axial current turns out to be:
\begin{equation*}
 \media{\h{j}^A_\mu}=-\frac{1}{3}\epsilon_{\mu\lambda\rho\sigma}\mathcal{S}^{\lambda,\rho\sigma} =
 2\iota\, w_\mu = W^A\, w_\mu,
\end{equation*}
where we used the (\ref{alphaw}). Hence, the eq.~(\ref{axcurr}) is recovered, with the 
same coefficient $W_A$ in (\ref{coeffw2}).
\footnote{While we were completing this work, a paper by Flachi and Fukushima appeared 
\cite{fukushima} where they calculated the axial current for a rotating system in curved 
spacetime. It can be shown that the small mass limit of (\ref{coeffw2}) corresponds to 
their result in flat spacetime.}

\section{Limiting cases}\label{limcase}

In this section we discuss some limiting cases which may be of interest for various physical 
situations. It is important to stress that all of these corrections, at least for free fields, 
are of quantum origin, and thus are expected to contribute in limiting cases where temperature 
is very low and/or the chemical potential stays finite, in presence of vorticity and acceleration.
Any coefficient $Y$ among those of eqs.~(\ref{boscoeff}),(\ref{fermcoeff}) for the stress-energy
tensor can be generally expressed as:
\begin{equation}\label{bigy}
\begin{split}
  Y=\frac{1}{\pi^2\beta^2}\int_0^\infty \frac{\di p}{E_p}  \left(n''_{F,B}(E_p-\mu)+ n''_{F,B}(E_p+\mu)\right)   
  (Ap^4+Bm^2 p^2+Cm^4),
\end{split}
\end{equation}
with $A$, $B$, $C$ some numerical constant and:
\begin{equation*}
\begin{split}
n''_{F,B}(E_p\pm\mu)=\frac{\di^2 }{\di E^2_p}\left(\frac{1}{\e^{|\beta|(E_p\pm\mu)}-\eta}
\right),
\end{split}
\end{equation*}
where $\eta$ is $+1$ for bosons $-1$ for fermions. Also, the axial current coefficient~(\ref{coeffw2}) 
can be recast in the form (\ref{bigy}) after integration by parts:
\begin{equation*}
W^A=\frac{1}{6 \pi^2 |\beta| }\int_0^\infty \frac{\di p}{E_p} \left(n_F''(E_p-\mu)+n_F''(E_p+\mu)\right) p^4.
\end{equation*}
%

\bigskip
\begin{center}
$\underline{\bm{m = 0}}$
\end{center}
\bigskip

The simplest case is the massless one, where the integral can be calculated analytically. For
the free massless Boson field the only physical chemical potential value is $\mu=0$, whereas, 
for fermions a non-vanishing $\mu$ is possible:
\begin{equation*}
    \int_0^\infty \di p \,p^k \left(\frac{1}{\e^{|\beta| (p-\mu)}+1}+
    \frac{1}{\e^{|\beta| (p-\mu)}+1}\right)=-|\beta|^{-k-1} k\, \Gamma(k)
    \left(\mathrm{Li}_{k+1}(-\e^{|\beta|\mu})+\mathrm{Li}_{k+1}(-\e^{-|\beta|\mu})\right)
\end{equation*}
where $\Gamma$ is the Euler gamma function and $\mathrm{Li}_{k}$ are the Polylogarithm function 
\cite{Abram}. The coefficients for massless Boson field at $\mu=0$ are reported in table~\ref{tab:Boson} 
and for massless Dirac field in table~\ref{tab:Fermion}.
\begin{table}[tb]
\small
\caption{The stress-energy tensor coefficients (\ref{coefficienti}) for a free Boson field: 
the first column reports the coefficients in the massless case with $\mu=0$, the second column 
reports the $n$th term of the expansion (\ref{yseries}), the third column reports the 
asymptotic expansion at low temperature~(\ref{lowtemp}). Our result for $W$ in the massless case 
agrees with that obtained in ref.~\cite{moore} for $\lambda_3$ (see eq.~(\ref{compar}))}.
\newcommand\bstrut{\vphantom{\displaystyle\frac{\pi^{2^T}}{\sum}}}
\label{tab:Boson}
\[
\begin{array}{|>{\displaystyle}l|>{\displaystyle}c|>{\displaystyle}c|>{\displaystyle}c|}
\hline
\vphantom{\frac{1}{2}} &m=0\, \mu=0& a_n &f(x)\quad x =m/T\gg 1\\ \hline
\bstrut \rho &
\frac{\pi^2}{15}T^4  &
\left[-(n x)^{-2 }K_2(n x)+(n x )^{-1}K_3(n x )  \right] & 
(1+27/8x)
\\
\bstrut p & \frac{\pi^2}{45}T^4 &
\left[(n x)^{-2 }K_2(n x)  \right]
&
\frac{1}{x}(1+15/8x)
\\
\bstrut U_\alpha &
\frac{(1-6\xi)}{6}T^4
&
 \f{1}{24 } \left[ ( n^2+24\xi x^{-2})K_2 (nx)+3(1-8\xi)nx^{-1} K_3( n x )\right]
&\frac{x}{24}(1+(39/8-24 \xi)/x)\\
\bstrut D_\alpha & \f{(6\xi-1)}{9}T^4 
& 
\f{1}{24 }\left[(12-48\xi )x^{-2} K_2(n x )+(24\xi -5) x^{-1}n K_3(n x )\right]&
\frac{1}{24}(24\xi-5+(456\xi-79)/8x)\\
\bstrut A & \f{(1-6\xi)}{6}T^4  &\f{1}{4 }\left[(4\xi-2 )x^{-2} K_2(n x )+(1-4\xi) x^{-1}n K_3(n x )\right] & 
\frac{1}{4}(1-4\xi+(19-108\xi)/8x)
\\
\bstrut U_w & \f{(1-4\xi)}{6}T^4
&
\f{1}{2 }\left[(1-4\xi)x^{-2} K_2(n x) \right]
&  \frac{1-4\xi}{2x}(1+15/8x)\\
\bstrut D_w & \f{\xi}{3}T^4 
&
\xi x^{-2}  K_2(n x )
&\frac{\xi}{x}(1+15/8x)\\
\bstrut W &\frac{(2\xi -1)}{6}T^4& \f{1}{2 }x^{-2}(2\xi-1)  K_2(n x )& \frac{2\xi-1}{2x}(1+15/8x)\\
\bstrut G & \f{(1+6\xi)}{18}T^4  
&
 \f{1}{6 } \left[(6\xi-3)x^{-2} K_2 (n x ) +n x^{-1} K_3( nx )\right]
& \frac{1}{6}(1+(6\xi+11/8)/x) \bigstrut[b] \\
\hline
\end{array}
\]
\end{table}
\begin{table}[tb]
\small
\caption{The stress-energy tensor coefficients (\ref{coefficienti}) for a free Dirac field: 
the first column reports the coefficients in the massless case, the second column 
reports the coefficient of the $n$th term of the expansion (\ref{yseries}), the third 
column reports the asymptotic expansion at low temperature~(\ref{lowtemp}).}
\label{tab:Fermion}
\newcommand\bstrut{\vphantom{\displaystyle\frac{\pi^{2^T}}{\sum}}}
\[
\begin{array}{|>{\displaystyle}l|>{\displaystyle}c|>{\displaystyle}c|>{\displaystyle}c|}
\hline
\vphantom{\frac12}&m=0& a_n & f(x)\quad x =m/T\gg 1
\\\hline
\vphantom{\sum^{\Lambda^2}_{\Lambda^2}}\rho& \left(\frac{7\pi^2}{60}+\frac{\mu^2}{2T^2}+\frac{\mu^4}{4\pi^2T^4}\right)T^4
&
 \left[(n x)^{-1}K_3( n x )  - (n x)^{-2} K_2 (n x )\right]
&  (1+27/8x)\\
\vphantom{\sum^{\Lambda^2}_{\Lambda^2}} p & \frac{1}{3}\left(\frac{7\pi^2}{60}+\frac{\mu^2}{2T^2}+\frac{\mu^4}{4\pi^2T^4}\right)T^4
&
(n x)^{-2}K_2(n x)
&  \frac{1}{ x}(1+15/8x)\\
\bstrut U_\alpha &
\frac{1}{24}\left(1+\frac{3 \mu^2}{\pi^2 T^2}\right)T^4
&
\f{1}{24} \left[ (3x^{-2}+n^2)  K_2( n x )\right]
&\frac{x}{24}(1+15/8x)\\
\bstrut D_\alpha & \f{1}{72} \left(1+\frac{3\mu^2}{\pi^2 T^2}\right)T^4
& 
\f{1}{24}\left[-3 x^{-2} K_2(n x )+n x^{-1} K_3(n x )\right]
&\frac{1}{24}(1+11/8x)\\
\vphantom{\sum}A & 0& 0& 0\\
\bstrut U_w & \f{1}{8} \left(1+\frac{3 \mu^2}{\pi^2 T^2}\right)T^4
&
\f{1}{8}\left[-x^{-2}  K_2(n x) +nx^{-1} K_3 (n x )\right]
& \frac{1}{8}(1+27/8x)\\
\bstrut D_w & \f{1}{24} \left(1+\frac{3 \mu^2}{\pi^2 T^2}\right)T^4
&
\f{1}{8}x^{-2}  K_2(n x )
&\frac{1}{8 x}(1+15/8x)\\
\vphantom{\sum} W &0& 0& 0\\
\vphantom{\sum^{\Lambda^2}_{\Lambda^2}} G & \f{1}{18}  \left(1+\frac{3 \mu^2}{\pi^2 T^2}\right)T^4
&
\f{1}{24} \left[n x^{-1}  K_3 (n x ) \right]
&\frac{1}{24}(1+35/8x)\\
\bstrut W\apic{A} & \f{T^3}{6}+ \frac{T\mu^2}{2\pi^2}
&
\f{1}{2m}\left[x^{-1}  K_0(n x )+2 n^{-1} x^{-2} K_1(n x )\right]
&\frac{1}{2m}(1+15/8x)\\
\hline
\end{array}
\]
\end{table}

\bigskip
\begin{center}
$\underline{\bm{m > |\mu| > 0}}$
\end{center}
\bigskip

For relativistic massive fields, one has two cases: $|\mu| < m$ and $|\mu|> m$ \footnote{The 
relativistic chemical potential $\mu$ is related to the one used in non-relativistic statistical
mechanics $\mu_{NR}$ by $\mu = \mu_{NR} + m$}. For $|\mu| < m$ it is possible 
to expand the distribution function:
\begin{equation*}
n''_{F,B}(E_p\pm\mu)=\sum_{n =1}^{\infty}\eta^{n+1}(-n|\beta|)^2 \e^{-|\beta|(n E_p\pm\mu)},
\end{equation*}
because $|\beta|(E_p\pm\mu) > 0$. Under this condition 
the distribution function $n_{F,B}$ can be expressed as a geometric series of the Boltzmann one.
Introducing this expansion, changing the integration variable to the rapidity $y$ with 
$p=m \sinh y$, defining $x = m/T = m |\beta|$, the eq.~(\ref{bigy}) can be rewritten as:
\begin{equation*}
\begin{split}
 Y=\frac{m^4}{ \pi^2}\sum_{n=1 }^\infty \eta^{n+1} 2 n^2 \cosh (n|\beta| \mu) \int_0^\infty \di y \, 
 \e^{- n x \cosh y}  (A\sinh^4 y+ B\sinh^2 y +C).
\end{split}
\end{equation*}
The integration can be carried out by using the integral representation of the modified Bessel 
function of the second kind, or McDonald functions $K_\nu(x)$ \cite{Abram}:
\begin{equation}\label{yseries}
  Y=(2S+1)\frac{m^4}{\pi^2}\sum_{n=1}^{\infty}\eta^{n+1} a_n(x) \cosh(n |\beta| \mu ),
\end{equation}
where the coefficients $a_n$ for bosons ($S=0,\,\eta=1$) are shown in table~\ref{tab:Boson} and for
fermions ($S=1/2,\,\eta=-1$) in table~\ref{tab:Fermion}. 

The above series is well suited to study the non-relativistic limit of the coefficients,
what happens when the mass is much larger than the temperature, that is $x \gg 1$. So, by using 
the asymptotic expansion of the McDonald functions \cite{Abram}:
\begin{equation*}
    K_\nu(n x)  \simeq \e^{-n x }\sqrt{\frac{\pi}{2 x}}\left[1+\frac{4\nu^2-1}{8 nx }+
    \frac{(4\nu^2-1)(4\nu^2-9)}{2!(8 nx)^2 }+\cdots\right].
\end{equation*}
In this limit $m \gg T$ either the particle or antiparticle contribution can be neglected, and
since $|\mu| < m$, the first term in the series (\ref{yseries}) is the dominant one and 
quantum statistics reduces to the Boltzmann limit. Thus, one can write the coefficient (\ref{bigy}) 
in the non-relativistic limit with $|\mu| < m$ for, e.g. particles with $\mu \ge 0$, to a
very good approximation as:
\begin{equation}\label{lowtemp}
Y=\f{(2S+1)m^4}{2 \sqrt{2}\, \pi ^{3/2} x^{3/2} }\e^{-|\beta| (m-\mu)}f\left(x\right)
\end{equation}
where $f $ is a polynomial of $1/x=T/m$ and its first leading terms are shown in 
tables~\ref{tab:Boson} and~\ref{tab:Fermion}. 

Note that eq.~(\ref{lowtemp}) can be rewritten as:
$$
Y= m \f{\di N}{\di^3 \x} f(x)
$$
where $\di N/\di^3 \x$ is the classical expression of particle density in the Boltzmann limit. 
This has an important consequence, that is all coefficients of the stress-energy tensor in 
eqs.~(\ref{boscoeff}) and (\ref{fermcoeff}) have a finite classical limit whose leading term is proportional to either 
mass times particle density or temperature times density. Therefore, the second-order corrections 
to the ideal form of the stress-energy tensor appearing in eq.~(\ref{setgen}) at thermodynamic equilibrium 
must be quantum \cite{Becattini:2015nva}, as they vanish in the limit $\hbar \to 0$ according to 
the (\ref{vectmagn}); this explains why genuine quadratic terms in thermal vorticity are not 
found in the Boltzmann kinetic approach to the gradient expansion of the stress-energy tensor 
\cite{jaiswal}.

The same conclusion applies to the vector currents coefficients, see eqs.~(\ref{boscurrcoeff}) and 
(\ref{fermcurrcoeff}), because at the same conditions they can be expressed in a fashion similar to (\ref{lowtemp}). Since 
vector current is odd under charge conjugation, relevant coefficients are an odd function of $\mu$ and 
vanish at zero chemical potential. The general coefficient $Y_{\text{curr}}$ among those of 
eqs.~(\ref{boscurrcoeff}) and (\ref{fermcurrcoeff}) can be written as, for $|\mu| < m$:
\begin{equation}\label{bigycurr}
\begin{split}
Y_{\text{curr}}=(2S+1)\frac{m^3}{\pi^2}\sum_{n=1}^{\infty}\eta^{n+1}b_n \sinh(n |\beta| \mu ),
\end{split}
\end{equation}
where the hyperbolic sine (odd function of $\mu$) replaces hyperbolic cosine. The non-relativistic 
limit is very similar to that of (\ref{lowtemp}):
\begin{equation}\label{lowtempcurr}
Y_{\text{curr}}=\f{(2S+1)m^3}{2 \sqrt{2}\, \pi^{3/2} x^{3/2} } \e^{-|\beta| (m-\mu)} g(x)
= \f{\di N}{\di^3 \x} g(x).
\end{equation}
The coefficients (\ref{currcoeff}) for bosons are reported in table~\ref{tab:Bosoni corrente}, 
and for fermions in table~\ref{tab:Fermion corrente}.
\begin{table}[b!t]
\small
\caption{The vector current coefficients (\ref{currcoeff}) for an ideal boson field: the 
first column is the result at $m=0$ and $\mu=0$, the second column is the generic terms of 
the series expansion (\ref{bigycurr}), the third column is the asymptotic expansion 
for low temperature~(\ref{lowtempcurr}).}
\label{tab:Bosoni corrente}
\[
\begin{array}{|>{\displaystyle}l|>{\displaystyle}c|>{\displaystyle}c|>{\displaystyle}c|}
\hline
\vphantom{\frac12}&m=0\quad\mu=0& b_n & g(x) \quad x = m/T \gg 1
\\\hline
\vphantom{\frac12}n&0&  (n x)^{-1} K_2(n x)&(1+ 15/8x)\\
N_\alpha\apic{V} &  0 &\f{1}{24} n^2 K_1(n x)& \frac{x}{24}(1+3/8x)\\
N_w\apic{V} &0 &0&0\\
\vphantom{\frac{\sum}{\sum}}G\apic{V} &0&-\f{1}{6}  x^{-1}n K_2(nx) &-\frac{1}{6}( 1+15/8x)\\
\hline
\end{array}
\]
\end{table} 
\begin{table}[b!t]
\small
\caption{The vector current coefficients (\ref{currcoeff}) for an ideal Dirac field: the first 
column is the result at $m=0$ , the second column is the generic terms of the series expansion 
(\ref{bigycurr}), the third column is the asymptotic expansion for low temperature~(\ref{lowtempcurr}).}
\label{tab:Fermion corrente}
\newcommand\bstrut{\vphantom{\displaystyle\left(\frac{\mu^2}{\pi^2 T^2}\right)}}
\[
\begin{array}{|>{\displaystyle}l|>{\displaystyle}c|>{\displaystyle}c|>{\displaystyle}c|}
\hline
\vphantom{\frac12}&m=0& b_n & g(x) \quad (m/T\gg 1)
\\\hline
\vphantom{\left(\frac{\pi^{2^T}}{\sum}\right)} n& \frac{T^2\mu}{3}\left(1+\frac{\mu^2}{\pi^2 T^2}\right) & (n x)^{-1} K_2(n x) & (1+15/8x)\\
\bstrut N_\alpha\apic{V} &  \f{T^2\mu}{4\pi^2} & \f{n^2}{96} \left[K_1(n x) + 3 K_3(nx ) \right] & \frac{x}{24}(1+27/8x)\\
\bstrut N_w\apic{V} & \f{T^2\mu}{4\pi^2} & \f{1}{8}n x^{-1}K_2(n x) & \frac{1}{8}(1+15/8x)   \\
\bstrut G\apic{V} & \f{T^2\mu}{6\pi^2} & \f{1}{12}n x^{-1} K_2(nx) &  \frac{1}{12}(1+15/8x) \\ 
\hline
\end{array}
\]
\end{table}

\bigskip
\begin{center}
$\underline{\bm{|\mu| > m > 0}}$
\end{center}
\bigskip

As has been mentioned, the previous expansion is possible only when $|\mu|< m$. For the Boson 
gas, at fixed charge (or particle, in the non-relativistic limit) density, at some very low 
temperature $T \sim 0$ the chemical potential attains the limiting value $\mu=m$ ($\mu_{NR} = 0$ in the
non-relativistic framework), implying the onset of Bose-Einstein condensation. 

In the fermion case, at very low temperature, the case $|\mu| > m$ is the so-called degenerate case. 
Indeed, when $\mu > m$, the Fermi-Dirac distribution function at $T=0$ becomes a step function:
\begin{equation*}
\begin{split}
\lim_{T\to 0 }n_{F}(E_p-\mu)=\lim_{|\beta|\to +\infty  }\frac{1}{\e^{|\beta| (E_p-\mu)}+1}=\theta(\mu -E_p)
\end{split}
\end{equation*}
and the antiparticle contribution vanishes. The coefficients at zero temperature, in the degenerate 
case, can be expressed in terms of the parameter of $b =\mu/m$ ad they are shown in 
table \ref{tab:degferm}.

\begin{table}[tb]
\small
\caption{The coefficients of the energy momentum tensor (\ref{coefficienti}), the vector current (\ref{currcoeff}) and axial current (\ref{axcurr}) for the Dirac field at $T=0$ and finite chemical potential $\mu$, where $b=\mu/m$.}
\label{tab:degferm}
\newcommand\bstrut{\vphantom{\displaystyle \frac{\displaystyle\sum}{2}}}
\[
\begin{array}{|>{\displaystyle}l|>{\displaystyle}c|}
\hline
 \vphantom{\frac{\displaystyle\sum}{\sum}}\rho& \frac{\mu^4}{8\pi^2}b^{-4}\left[(2\,b^2-1)b\sqrt{b^2-1}-\log\left(b+\sqrt{b^2-1}\right)\right]  \\
 p&\frac{\mu^4}{24\pi^2}b^{-4}\left[(2\,b^2-5)b\sqrt{b^2-1}+3\log\left(b+\sqrt{b^2-1}\right)\right]  \\
 \bstrut\beta^2U_\alpha& \frac{\mu^2}{24 \pi^2  }\frac{b(3b^2-4)}{(b^2-1)^{3/2}}  \\
\bstrut\beta^2 D_\alpha& \frac{\mu^2}{24 \pi^2  }\frac{b}{\sqrt{b^2-1}}  \\
\bstrut\beta^2 U_w& \frac{\mu^2}{8 \pi^2  }\frac{3b^2-2}{b\sqrt{b^2-1}}  \\
 \bstrut\beta^2D_w& \frac{\mu^2}{8 \pi^2  }\frac{\sqrt{b^2-1}}{b} \\
\bstrut\beta^2 G&\frac{\mu^2}{24 \pi^2  }\frac{4b^2-3}{b\sqrt{b^2-1}} \\
 \bstrut n& \frac{\mu^3}{3 \pi^2 }\frac{(b^2-1)^{3/2}}{b^3}  \\
\bstrut\beta^2 N\apic{V}_\alpha& \frac{\mu}{24 \pi^2 }\frac{6b^4-9b^2+2}{b(b^2-1)^{3/2}}  \\
\bstrut\beta^2 N\apic{V}_w& \frac{\mu}{8 \pi^2  }\frac{2b^2-1}{b\sqrt{b^2-1}}  \\
\beta^2 G\apic{V} &\frac{\mu}{12 \pi^2  }\frac{2b^2-1}{b\sqrt{b^2-1}}  \\
|\beta| W\apic{A} & \vphantom{\frac{\displaystyle\sum}{\sum}}\frac{\mu^2}{2 \pi^2  }\frac{\sqrt{b^2-1}}{b}   \\
\hline
\end{array}
\]
\end{table}
Although all second-order coefficients vanish in the limit $T \to 0$, it is worth pointing out that  
a corresponding $1/T^2$ or $1/T$ factor appears in the quadratic terms in thermal vorticity, recalling 
that: 
\begin{equation*}
\begin{split}
\alpha^\mu = \frac{a^\mu}{T} \quad \quad w^\mu = \frac{\omega^\mu}{T}
\end{split}
\end{equation*}
as seen in Sect.~\ref{accvort}. Therefore, all quadratic corrections to the stress-energy tensor
in the (\ref{setgen}) remain finite in the zero temperature limit in acceleration and vorticity.
Particularly, all corrections to the stress-energy tensor, from table \ref{tab:degferm} turn out
to be of the form $\mu^2 a^2 F(b)$ or $\mu^2 \omega^2 F(b)$ where $F(b)$ is a function of $b=\mu/m$. 

In principle, these corrections might be phenomenologically relevant for very cold fermion stars,
if their magnitude was comparable to the ideal term $\mu^4$ of energy density and pressure (see 
table~\ref{tab:degferm}). However, the typical values of the baryon chemical potential (Fermi energy), 
spinning frequency and gravitational acceleration of a neutron star imply a tiny ratio $a/\mu \approx \omega/\mu 
\approx 10^{-19}\div 10^{-27}$ and the functions $F(b)$ remain finite even in the $b \to \infty$ 
limit, that is for a very light fermion. Therefore, these corrections, at least for a free field,
are negligible.

\medskip
\begin{center}
\line(1,0){250}
\end{center}
\medskip

In general, one can argue that these corrections might be relevant for very cold massive gases 
subject to large accelerations and rotations. Particularly, from table \ref{tab:Boson} and 
\ref{tab:Fermion} it can be realized that in the non-relativistic limit $m/T \gg 1$ the ratio
$D_w/p$ is of the order 1, so that at finite $a$ or $\omega$ in the $T \to 0$ limit the contribution
of the corrections to the pressure in the stress-energy tensor (\ref{setgen}) blows up. Obviously, 
when the ratios $a/T$ or $\omega/T$ are ${\cal O}(1)$ the whole expansion method breaks down,
but this behaviour points to a relevance of the quantum effects in the low $T$ limit for sufficiently large $a$ and $\omega$.

These corrections may also play a role in high energy nuclear collisions and Quark Gluon Plasma
(QGP) physics. The very recent measurement of the $\Lambda$ hyperon polarization with respect to 
the reaction plane indicates a magnitude of the thermal vorticity $\varpi$ at the hadronization 
stage of the order of $10^{-2}$ at a centre-of-mass energy ${\cal O}(10)$ GeV. As thermal vorticity 
is presumably much larger in the early stage of the QGP expansion, the second-order non-dissipative corrections may be of some relevance for the hydrodynamic evolution and could compete with the 
first-order dissipative terms. 

\section{Comparison with previous determinations}

As has been already mentioned, second order coefficients for the stress-energy tensor were classified 
in~\cite{baier,Romatschke:2009kr,Moore:2010bu} in the Landau frame. Following their notation,
for a non-conformal fluid in flat spacetime, the relevant coefficients are $\lambda_3,\,\lambda_4,\,
\xi_3,$ and $\xi_4$, dubbed as thermodynamical in~\cite{moore} because these terms survive 
at thermodynamic equilibrium with rotation or acceleration. 
Since we adopted the $\beta$-frame and not the Landau frame, before comparing the coefficients 
we have to change the hydrodynamic frame. Up to second order in vorticity the relation between 
the fluid four velocity in the Landau frame $u_L$ and in $\beta$-frame $u$ is found diagonalizing 
the stress-energy tensor expansion in $\beta$-frame~(\ref{setgen}), see Appendix \ref{appa}:
\begin{equation*}
  u_L = u + \frac{G}{\rho+p} \gamma+ \mathcal{O}(\varpi^3).
\end{equation*}
Moreover in~\cite{baier,Romatschke:2009kr,Moore:2010bu} a different definition of temperature 
and chemical potential is introduced, such that the energy and particle density are the same
functions of temperature and chemical potential as in homogeneous thermodynamic equilibrium 
$$ 
\rho_L\equiv T_{\mu\nu}u^\mu_L u^\nu_L=\rho(T',\mu') \qquad  n_L\equiv j_{\mu}u^\mu_L=n(T',\mu').
$$
The relations between $T,\mu$ and $T',\mu'$ are reported in eqs.~(\ref{lmc2}) and~(\ref{lmc3}).

The stress-energy tensor and current expansions at the second order in thermal vorticity 
in the Landau frame read (see Appendix~\ref{appa} for details):
\begin{equation*}
\begin{split}
T^{\mu\nu}&=\rho(T',\mu') u_L^\mu u_L^\nu-\left(p(T',\mu') -\alpha^2  D'_\alpha -w^2 D'_w \right)
\Delta_L^{\mu\nu} + A \alpha^\mu\alpha^\nu+ W w^\mu w^\nu+ \mathcal{O}(\varpi^3),
\\
j^\mu &=n(T',\mu') u_L^\mu+\left(G^V-\frac{n(T',\mu')}{\rho(T',\mu')+p(T',\mu')}G \right)\gamma^\mu+ \mathcal{O}(\varpi^3),
\end{split}
\end{equation*}
where the definitions of $D'_\alpha$ and $D'_w$ in terms of the $\beta$-frame coefficients are:
\begin{equation*}
\begin{split}
D'_\alpha=D_\alpha-U_\alpha
\frac{\frac{\partial p}{\partial T}\frac{\partial n}{\partial \mu}-\frac{\partial p}{\partial \mu}\frac{\partial n}{\partial T}}
{\frac{\partial \rho}{\partial T}\frac{\partial n}{\partial \mu}-\frac{\partial \rho}{\partial \mu}\frac{\partial n}{\partial T}}
-N_\alpha
\frac{\frac{\partial p}{\partial T}\frac{\partial \rho}{\partial \mu}-\frac{\partial p}{\partial \mu}\frac{\partial \rho}{\partial T}}
{\frac{\partial \rho}{\partial T}\frac{\partial n}{\partial \mu}-\frac{\partial \rho}{\partial \mu}\frac{\partial n}{\partial T}},
\\
D'_w=D_w-U_w\frac{\frac{\partial p}{\partial T}\frac{\partial n}{\partial \mu}-\frac{\partial p}{\partial \mu}\frac{\partial n}{\partial T}}
{\frac{\partial \rho}{\partial T}\frac{\partial n}{\partial \mu}-\frac{\partial \rho}{\partial \mu}\frac{\partial n}{\partial T}}
-N_w
\frac{\frac{\partial p}{\partial T}\frac{\partial \rho}{\partial \mu}-\frac{\partial p}{\partial \mu}\frac{\partial \rho}{\partial T}}
{\frac{\partial \rho}{\partial T}\frac{\partial n}{\partial \mu}-\frac{\partial \rho}{\partial \mu}\frac{\partial n}{\partial T}}.
\end{split}
\end{equation*}
A comparison of the above decomposition with the one in~\cite{baier,Romatschke:2009kr,Moore:2010bu}
allows to extract the relations between $\lambda_3,\,\lambda_4,\,\xi_3$ and $\xi_4$ and our set of 
independent coefficients $D'_\alpha,\,D'_w,\,A$ and $W$ \footnote{Please note that these transformation
relations differs from those quoted in ref.~\cite{Becattini:2015nva} as the temperature and chemical
potential redefinition was not taken into account therein.}.
\begin{align}\label{compar}
\frac{W}{T'^2}&=\lambda_3 &  \frac{A}{T'^2}&=9\lambda_4\nonumber\\
\frac{D'_w}{T'^2}&=\left(2\xi_3-\frac{\lambda_3}{3}\right) &
\frac{D'_\alpha}{T'^2}&=\left(9\xi_4-3\lambda_4\right).
\end{align}
The above equalities can be inverted to give:
\begin{align}\label{compar2}
\lambda_3 &=\frac{W}{T'^2} &  \lambda_4 &= \frac{1}{9}\frac{A}{T'^2} \nonumber\\
\xi_3 &= \frac{1}{2} \frac{D'_w +W/3}{T'^2} & \xi_4 &= \frac{1}{9} \frac{D'_\alpha+A/3}{T'^2}.
\end{align}

The coefficients $\lambda_3$ and $\lambda_4$ were reported in refs.~\cite{moore} and ref.~\cite{megias}
for free massless bosons and fermions. Our result of $\lambda_3 = W/T^2$
for bosons reported in the first column of table~\ref{tab:Boson} agrees with that quoted in ref.~\cite{moore} for a stress-energy tensor with $\xi = 0$.  Furthermore, in ref.~\cite{moore} $\lambda_4$ is argued to be zero in conformal case and correspond to our result for the improved stress-energy tensor $\xi=1/6$. Then, our results of $\lambda_3 = W/T^2$ for massless fermions reported in the first column of table~\ref{tab:Fermion} 
-both vanishing - agree with the results quoted in ref.~\cite{megias} whereas they are in 
disagreement with those quoted in ref.~\cite{moore} at $\mu=0$, equal to $\lambda_3=T^2/12$.

\section{Conclusions}

In conclusion, we have studied quantum relativistic free fields of spin 0 and $1/2$ at general 
thermodynamic equilibrium with non-vanishing acceleration and vorticity and we have calculated
the thermodynamic coefficients of a second-order expansion of the stress-energy tensor in thermal 
vorticity tensor, which includes acceleration and vorticity vectors, also with a finite value of the chemical
potential. We have also determined the leading order coefficients for the vector and axial currents. 
Such corrections may be phenomenologically relevant for system with very high acceleration, 
or vorticity as in the early stage of relativistic heavy ion collisions \cite{star}.

We have shown, like in ref.~\cite{Becattini:2015nva}, that our method is very convenient to
determine the coefficient of these non-dissipative (i.e. persisting in thermodynamic equilibrium)
terms involving vorticity and acceleration, envisaged in the general hydrodynamic expansion of 
the stress-energy tensor. We have extended the results of our previous work and we have 
compared our results with the definitions used in the so-called Landau frame. We reinforce our 
previous conclusion \cite{Becattini:2015nva} that these terms are of quantum nature. 

We have studied the relation between axial current and vorticity known as Axial Vortical Effect 
(\ref{axcurr}) for the free Dirac field. 
The coefficient found for the massless 
field (\ref{AVEnomass}), which, in our calculation, is simply an effect of rotation at equilibrium, 
coincides with those quoted in literature and attributed to the gauge and gravitational
anomalies as the pertaining Kubo formulae are identical. We cannot, for the present, demonstrate 
that the two derivations are equivalent.

\acknowledgments

Useful discussions with K. Fukushima, A. Jaiswal and P. Kovtun are gratefyully acknowledged.
E. Grossi was supported by the DFG Collaborative Research Centre “SFB 1225 (ISOQUANT)”
and carried out most of his work as a post-doctoral fellow in the University
of Florence supported by the Ente Cassa di Risparmio di Firenze, grant no. 20140754.
This work was partly supported by the University of Florence grant {\em Fisica dei plasmi 
relativistici: teoria e applicazioni moderne}.

\appendix

\section{Landau frame}\label{appa}

In this work we have used the so-called $\beta$ \cite{becalocal} or thermodynamic \cite{kovtun,Van:2013sma} frame,
determined by the eq.~(\ref{fourv}) where $\beta$ is the four-temperature fulfilling the 
Killing equation~(\ref{killing}). It has been  shown in ref.~\cite{becalocal} that this 
frame does not coincide with Landau's frame, where the four-velocity $u_L$ is defined by the 
timelike eigenvector of the stress-energy tensor, in thermodynamic equilibrium
situations where the thermal vorticity is non-vanishing. This is also apparent from the
expansion of the stress-energy tensor at the second order in thermal vorticity, eq.~(\ref{setgen})
that is rewritten here for the sake of clarity:
$$
T^{\mu\nu}=(\rho-\alpha^2 U_\alpha -w^2 U_w)u^\mu u^\nu-(p-\alpha^2D_\alpha-w^2D_w)
\Delta^{\mu\nu}+A\alpha^\mu\alpha^\nu+Ww^\mu w^\nu+ G(u^\mu\gamma^\nu+u^\nu\gamma^\mu)+\mathcal{O}(\varpi^2).
$$
The term involving the $\gamma$ four-vector makes the four-velocity $u$, defined in eq.~(\ref{fourv})
no longer an eigenvector of $T^{\mu\nu}$, hence not the Landau frame velocity $u_L$.
Of course, it is possible to change the frame definition and rewrite the stress-energy
tensor expression (\ref{setgen}) in the new frame, e.g. the Landau frame. This entails
a transformation rule for the second-order coefficients $U_\alpha,\,A,\ldots$ as well.

The transformation to the Landau frame requires the diagonalization of the energy momentum tensor
and the determination of its unique time-like eigenvector $u_L$:
$$
u_{L\mu} T^{\mu\nu}=\rho_L u_L^\nu
$$
where $\rho_L$ is the eigenvalue, that is the proper energy density in the Landau frame.
By looking at the stress-energy tensor expression (\ref{setgen}) it can be readily realized
that the eigenvector $u_{L}$ ought to be a linear combination of the $u$ (the $\beta$ frame
velocity) and  $\gamma$:
\be\label{landau}
u_{L\mu}=a\, u_{\mu} + b\, \frac{\gamma^\mu}{|\gamma|},
\ee
where $|\gamma| \equiv \sqrt{-\gamma^2}$ and $a$ and $b$ two unknown constants 
such that $a^2 - b^2 = 1$. Contracting the eq.~(\ref{setgen}) with $u_L$ and using
(\ref{landau}),(\ref{setgen}) one obtains:
\begin{equation*}
u_{L\mu} T^{\mu\nu} =a \rho_L u^\mu + b \rho_L\frac{\gamma^\mu}{|\gamma|} = 
a(\rho-\alpha^2 U_\alpha -w^2 U_w) u^\nu+a G\gamma^\nu -b(p-\alpha^2D_\alpha-w^2D_w) \frac{\gamma^\mu}{|\gamma|}-b G |\gamma| u^\nu
\end{equation*} 
which implies:
\begin{equation}\label{landsol}
\begin{split}
a \rho_{\rm eff} -b G |\gamma| =a \rho_L,\\
a G |\gamma| - b p_{\rm eff}= b \rho_L ,
\end{split}
\end{equation}
where $\rho_{\rm eff}=\rho-\alpha^2 U_\alpha - w^2 U_w$ and $p_{\rm eff}
=p-\alpha^2D_\alpha-w^2D_w$.

One can algebraically solve the equations (\ref{landsol}) to determine $a, b, \rho_L$ taking the 
constraint $a^2-b^2=1$ into account. However, since we are dealing with an expansion of
the stress-energy tensor to second order in $\varpi$, it is sufficient and more convenient
to find an approximate solution of the equations at the same order in this parameter. This
can be done by observing that the parameter $b$ must be ``small'' as for $\varpi \to 0$ 
one expects Landau and $\beta$ frame to coincide in eq.~(\ref{landau}). Therefore, from
the first of (\ref{landsol}), one has:
$$
  \rho_L \simeq \rho_{\rm eff}
$$  
and, from the second:
$$
  b \simeq \frac{G |\gamma|}{\rho_{\rm eff}+ p_{\rm eff}} \simeq  \frac{G |\gamma|}{\rho+p}
$$
where we have kept only second-order terms in $\varpi$ keeping in mind that $\gamma = {\cal O}(\varpi^2)$. 
At the same order of approximation, the coefficient $a \simeq 1$.

The eq.~(\ref{landau}) relating the Landau and $\beta$ frames is thus:
\be\label{landau2}
  u_L = u + \frac{G}{\rho+p} \gamma + {\cal O}(\varpi^3)
\ee
and the stress-energy tensor in the Landau frame at the second order in thermal vorticity
reads:
\begin{equation}\label{setlandau}
\begin{split}
 T^{\mu\nu}=(\rho-\alpha^2 U_\alpha -w^2 U_w) u_L^\mu u_L^\nu-(p-\alpha^2D_\alpha-w^2D_w)\Delta_L^{\mu\nu}+
 A \alpha^\mu \alpha^\nu + W w^\mu w^\nu.
\end{split}
\end{equation}
We can also write the conserved vector current (\ref{currdec}) in the Landau frame by using (\ref{landau2}):
\begin{equation}\label{landcurr}
\begin{split}
j^\mu(x) &=n u^\mu  -(\alpha^2 N_\alpha+ w^2 N_w )u^\mu +G^V\gamma^\mu  \\
 &=n u_L^\mu  -(\alpha^2 N_\alpha+ w^2 N_w )u_L^\mu +\left( G^V-n \frac{G}{\rho+p} \right)\gamma^\mu
 + {\cal O}(\varpi^3).
\end{split}
\end{equation}

It has become customary in the literature to include in the specification of the Landau frame
a redefinition of the temperature and the chemical potential in order to avoid corrections to 
the equilibrium energy and charge density \cite{kovtun,minwalla}:
\begin{equation}\label{lmc}
   \rho_L = \rho(T',\mu'),  \qquad \qquad n_L = n(T',\mu').
\end{equation}
The possibility to redefine the temperature is usually advocated in out-of-equilibrium situations,
and yet its application in global equilibrium situation with $\varpi\ne 0$ is conceptually very 
questionable (see also discussions in refs.~\cite{Becattini:2015nva,becalocal}) because it deprives 
temperature one of its key relativistic features - crucial to define equilibrium - that is being 
the inverse of the magnitude of a Killing vector field. If one pursues the implementation of (\ref{lmc}) 
anyway, 
\begin{equation}\label{lmc2}
\begin{split}
\rho(T',\mu') = \rho(T,\mu)-\alpha^2 U_\alpha -w^2 U_w,\\
 n(T',\mu')   =  n(T,\mu) - \alpha^2 N_\alpha -w^2 N_w,
\end{split}
\end{equation}
where we have used the previous results, that is $\rho_L = \rho_{\rm eff}$ and the eq.~(\ref{landcurr}).
In order to find the $T'$ and $\mu'$ we can perform a Taylor expansion of the temperature 
and chemical potential in powers of $\alpha^2$ and $w^2$:
\begin{equation}\label{lmc3}
 \begin{split}
   T'= T+ \frac{\partial T'}{\partial \alpha^2} \alpha^2+ \frac{\partial T'}{\partial w^2} w^2,\\
  \mu' =\mu + \frac{\partial \mu'}{\partial \alpha^2} \alpha^2+ \frac{\partial \mu'}{\partial w^2} w^2,
\end{split}
\end{equation}
which, once plugged into the (\ref{lmc2}) yield, after a Taylor expansion at the second order in 
$\varpi$:
\begin{equation*}
\begin{split}
\frac{\partial \rho}{\partial T} \left( \frac{\partial T'}{\partial \alpha^2} \alpha^2+ 
\frac{\partial T'}{\partial w^2} w^2 \right)+ \frac{\partial \rho}{\partial \mu} \left(
\frac{\partial \mu'}{\partial \alpha^2} \alpha^2+ \frac{\partial \mu'}{\partial w^2} w^2 \right) 
= -\alpha^2 U_\alpha -w^2 U_w;
\\ 
\frac{\partial n}{\partial T} \left( \frac{\partial T'}{\partial \alpha^2} \alpha^2+ 
\frac{\partial T'}{\partial w^2} w^2 \right)+ \frac{\partial n}{\partial \mu} \left(
\frac{\partial \mu'}{\partial \alpha^2}\alpha^2+ \frac{\partial \mu'}{\partial w^2} w^2\right)
= -\alpha^2 N_\alpha - w^2 N_w.
\end{split}
\end{equation*}
Equating the coefficients of $\alpha^2$ and $w^2$ on both sides, we obtain the solution:
\begin{equation*}
\begin{split}
\frac{\partial T'}{\partial \alpha^2} =\frac{-U_\alpha\frac{\partial n}{\partial \mu}+N_\alpha\frac{\partial \rho}{\partial \mu}}
{\frac{\partial \rho}{\partial T}\frac{\partial n}{\partial \mu}-\frac{\partial \rho}{\partial \mu}\frac{\partial n}{\partial T}},
\quad \quad 
\frac{\partial T'}{\partial w^2} =\frac{-U_w\frac{\partial n}{\partial \mu}+N_w\frac{\partial \rho}{\partial \mu}}
{\frac{\partial \rho}{\partial T}\frac{\partial n}{\partial \mu}-\frac{\partial \rho}{\partial \mu}\frac{\partial n}{\partial T}},\\
\frac{\partial \mu'}{\partial \alpha^2} =\frac{-N_\alpha\frac{\partial \rho}{\partial T}+U_\alpha\frac{\partial n}{\partial T}}
{\frac{\partial \rho}{\partial T}\frac{\partial n}{\partial \mu}-\frac{\partial \rho}{\partial \mu}\frac{\partial n}{\partial T}},\quad \quad 
\frac{\partial \mu'}{\partial w^2} =\frac{-N_w\frac{\partial \rho}{\partial T}+U_w\frac{\partial n}{\partial T}}
{\frac{\partial \rho}{\partial T}\frac{\partial n}{\partial \mu}-\frac{\partial \rho}{\partial \mu}\frac{\partial n}{\partial T}}.
\end{split}
\end{equation*}
We can replace these derivatives into the (\ref{lmc3}) to obtain the relation between the
proper temperature and chemical potential and the new $T'$ and $\mu'$. These relations 
allow to express all the thermodynamic functions with the new arguments. Clearly, we can
neglect any term beyond the second order in $\varpi$. Particularly, in the stress-energy
tensor expression at the second order (\ref{setlandau}), the only relevant thermodynamic
function which gets modified is the one involving pressure, 
\begin{equation*}
 \begin{split}
p(T,\mu)=p(T',\mu')-\alpha^2\left(\frac{\partial p}{\partial T}\frac{\partial T}{\partial \alpha^2}
+\frac{\partial p}{\partial \mu}\frac{\partial \mu}{\partial \alpha^2}\right)-
w^2\left(\frac{\partial p}{\partial T}\frac{\partial T}{\partial w^2}+\frac{\partial p}{\partial \mu}
\frac{\partial \mu}{\partial w^2}\right);
\end{split}
\end{equation*}
so that (\ref{setlandau}) and (\ref{landcurr}) become:
\begin{equation*}
\begin{split}
T^{\mu\nu}&=\rho(T',\mu') u_L^\mu u_L^\nu-\left(p(T',\mu') -\alpha^2  D'_\alpha -w^2 D'_w \right)
\Delta_L^{\mu\nu} + A \alpha^\mu\alpha^\nu+ W w^\mu w^\nu+ \mathcal{O}(\varpi^3),
\\
j^\mu &=n u_L^\mu+\left(G^V-\frac{n}{\rho+p}G \right)\gamma^\mu+ \mathcal{O}(\varpi^3),
\end{split}
\end{equation*}
where $D'_\alpha$ and $D'_w$ read:
\begin{equation*}
\begin{split}
D'_\alpha=D_\alpha-U_\alpha
\frac{\frac{\partial p}{\partial T}\frac{\partial n}{\partial \mu}-\frac{\partial p}{\partial \mu}\frac{\partial n}{\partial T}}
{\frac{\partial \rho}{\partial T}\frac{\partial n}{\partial \mu}-\frac{\partial \rho}{\partial \mu}\frac{\partial n}{\partial T}}
+N_\alpha
\frac{\frac{\partial p}{\partial T}\frac{\partial \rho}{\partial \mu}-\frac{\partial p}{\partial \mu}\frac{\partial \rho}{\partial T}}
{\frac{\partial \rho}{\partial T}\frac{\partial n}{\partial \mu}-\frac{\partial \rho}{\partial \mu}\frac{\partial n}{\partial T}},
\\
D'_w=D_w-U_w\frac{\frac{\partial p}{\partial T}\frac{\partial n}{\partial \mu}-\frac{\partial p}{\partial \mu}\frac{\partial n}{\partial T}}
{\frac{\partial \rho}{\partial T}\frac{\partial n}{\partial \mu}-\frac{\partial \rho}{\partial \mu}\frac{\partial n}{\partial T}}
+N_w
\frac{\frac{\partial p}{\partial T}\frac{\partial \rho}{\partial \mu}-\frac{\partial p}{\partial \mu}\frac{\partial \rho}{\partial T}}
{\frac{\partial \rho}{\partial T}\frac{\partial n}{\partial \mu}-\frac{\partial \rho}{\partial \mu}\frac{\partial n}{\partial T}}.
\end{split}
\end{equation*}
%

\section{Relations between coefficients}\label{appb}

Herein we derive the relations between coefficients (\ref{coeffrel}) enforcing the continuity
equation for the mean value of the stress-energy tensor $T_{\mu\nu}$ at second order in the
thermal vorticity $\varpi$
\begin{equation}\label{setapp}
\begin{split}
 T_{\mu\nu}= & (\rho-\alpha^2U_\alpha-w^2U_w)u_\mu u_\nu
      -(p-\alpha^2 D_\alpha-w^2D_w)\Delta_{\mu\nu}
       +A\alpha_\mu\alpha_\nu+Ww_\mu w_\nu +G ( u_\mu \gamma_\nu + u_\nu \gamma_\mu).
\end{split}
\end{equation}
Firstly, we observe that scalar thermodynamic functions depend on spacetime coordinates 
only through the magnitude of the four-temperature $F(|\beta|)=F(\sqrt{\beta^2(x)})$, thus:
\begin{equation}\label{gradscalar}
 \de_\nu F(|\beta|)=\f{\de|\beta|}{\de x^\nu} \f{\de F(|\beta|)}{\de|\beta|}=
\f{1}{2|\beta|}\de_\nu\big(\beta^\lambda\beta_\lambda\big)\f{\de F(|\beta|)}{\de |\beta|}
 =\f{1}{|\beta|}\beta^\lambda\de_\nu\beta_\lambda \f{F(|\beta|)}{\de |\beta|}
 =-u^\lambda\varpi_{\nu\lambda}\f{\de F(|\beta|)}{\de |\beta|} =-\alpha_\nu\f{\de F(|\beta|)}{\de |\beta|}
\end{equation}
where we have used the definition of $u$ in eq.~(\ref{fourv}), the eqs.~(\ref{thvort}) 
and (\ref{alphaw}).

We start by reckoning the gradients of the four-vectors $\{u,\alpha,w,\gamma\}$ which are
needed to calculate the stress-energy tensor divergence. First, we observe that, because
of the Killing equation (\ref{killing}), and using the eqs.~(\ref{fourv}), (\ref{thvort}) 
and (\ref{alphaw}):
$$
 0 = \beta^\mu (\partial_\mu \beta_\nu + \partial_\nu \beta_\mu) = |\beta| u^\mu
 \varpi_{\nu\mu} + \frac{1}{2} \partial_\nu {\beta^2} = |\beta| \alpha_\nu + 
 \frac{1}{2} \partial_\nu {\beta^2} 
$$
so that:
\be\label{alpha2}
 \alpha_\nu = - \frac{1}{2|\beta|} \partial_\nu {\beta^2}
\ee
and, contracting with $u$ again:
\be\label{dbeta}
   u^\nu \partial_\nu \beta^2  = D \beta^2 = 0 
\ee
what we already saw in section (\ref{accvort}). This equation implies that any scalar
function, whose argument are $\beta^2$ and $\xi$, has a vanishing derivative along the
flow, that is:
\be\label{df0}
   D F(\beta^2,\xi) = 0.
\ee
Now, we can find the gradient of $u$ as defined by the eq.~(\ref{fourv}):
\begin{equation}\label{gradu}
 \de_\nu u_\mu= \f{\de_\nu\beta_\mu}{|\beta|}+\beta_\mu\de_\nu\Big(\f{1}{|\beta|}\Big)=
 \f{1}{|\beta|}\big(\varpi_{\mu\nu}+\alpha_\nu u_\mu\big),
\end{equation}
where we have used again the eq.~(\ref{fourv}) and the eq.~(\ref{alpha2}). The divergence of $u$ 
then is:
\begin{equation}\label{divu}
 \de_\mu u^\mu = \de_\mu \frac{\beta^\mu}{|\beta|} = 0
\end{equation}
because of the Killing vector equation (\ref{killing}), which obviously imply 
$\partial_\mu \beta^\mu = 0$ and the (\ref{dbeta}).
Instead the derivative along its direction is
\begin{equation}\label{comderu}
 u_\rho\de^\rho u_\mu=\f{u_\rho}{|\beta|}\big(\varpi_\mu^{\,\rho}+\alpha^\rho u_\beta\big)=\f{\varpi_{\mu\rho}u^\rho}{|\beta|}=\f{\alpha_\mu}{|\beta|}.
\end{equation}
Then, let us calculate the derivative of $\alpha$, keeping in mind that $\varpi$ is constant:
\begin{equation}\label{gradalpha}
 \de_\mu\alpha_\nu=\varpi_{\nu\rho}\de_\mu u^\rho=\f{\varpi_{\nu\rho}}{|\beta|}\big( \varpi^\rho_{\phantom{\mu}\mu}+\alpha_\mu u^\rho\big)=
 \f{1}{|\beta|}\big(\varpi_{\nu\rho}\varpi^\rho_{\phantom{\mu}\mu}+\alpha_\mu\alpha_\nu\big);
\end{equation}
using the thermal vorticity decomposition~(\ref{vortdec}) and the Levi-Civita tensor properties 
we can express the previous formula in terms of the tetrad vectors
\begin{equation}\label{gradalpha2}
 \de_\mu\alpha_\nu=\f{1}{|\beta|}\left(-w_\mu w_\nu+\gamma_\mu u_\nu-u_\mu\gamma_\nu
 -(w^2+\alpha^2)u_\mu u_\nu+g_{\mu\nu}w^2\right),
\end{equation}
whence we obtain the divergence:
\begin{equation*}
 \de \cdot \alpha =\f{1}{|\beta|}\left(w^2-(w^2+\alpha^2)+4w^2\right)
 =\f{1}{|\beta|}\big( 2w^2-\alpha^2\big),
\end{equation*}
as well as the gradient of $\alpha^2$:
\be\label{dalpha2}
  \partial_\mu \alpha^2 = 2 \alpha^\nu \partial_\mu \alpha_\nu = 
  \f{2}{|\beta|} ( -\alpha \cdot w\, w_\mu + w^2 \alpha_\mu) 
\ee
taking into account that $\alpha \cdot u = \alpha \cdot \gamma = 0$.

Likewise, we can calculate the derivative of $w$ by using its definition (\ref{alphaw}) and
the (\ref{gradu}):
$$
 \de_\mu w_\nu=-\f{1}{2}\epsilon_{\nu\rho\sigma\lambda}\varpi^{\rho\sigma}\de_\mu u^\lambda
 =\f{-1}{2|\beta|}\epsilon_{\nu\rho\sigma\lambda}\varpi^{\rho\sigma}\varpi^\lambda_{\phantom{\mu}\mu}+\f{\alpha_\mu w_\nu}{|\beta|}\,.
$$
Replacing the eq.~(\ref{vortdec}) in the first term of the right hand side and using
the properties of the Levi-Civita tensor, it can be shown that:
\begin{equation}\label{dw}
\de_\mu w_\nu = \f{1}{|\beta|} \left( - g_{\mu\nu} \alpha \cdot w  + \alpha_\mu w_\nu \right)
\end{equation}
so that its divergence is
\begin{equation}\label{divw}
 \de_\mu w^\mu =-\f{3}{|\beta|}(w\cdot\alpha) 
\end{equation}
and the gradient of $w^2$:
\be\label{dw2}
  \partial_\mu w^2 = 2 w^\nu \partial_\mu w_\nu = 
  \f{2}{|\beta|} ( -\alpha \cdot w\, w_\mu + w^2 \alpha_\mu) = \partial_\mu \alpha^2.
\ee

In order to calculate the derivative of the last relevant vector field $\gamma$ (\ref{gamma}), 
one can first show that, by using \mbox{$\Delta_{\lambda\mu}=g_{\lambda\mu}-u_\lambda u_\mu$} 
and (\ref{vortdec}), it can be expressed as:
$$
 \gamma^\mu =\alpha^\rho\varpi_{\rho\lambda}\Delta^{\lambda\mu}
 =(\alpha\cdot\varpi)^\mu-\varpi_{\rho\lambda}\alpha^\rho u^\lambda u^\mu
 =(\alpha\cdot\varpi)^\mu-(\alpha_\rho u_\lambda-\alpha_\lambda u_\rho)\alpha^\rho 
 u^\lambda u^\mu=(\alpha\cdot\varpi)^\mu-\alpha^2 u^\mu
$$
so that its divergence vanishes
$$
 \de_\mu \gamma^\mu  =\de_\mu\big[(\alpha\cdot\varpi)^\mu-\alpha^2 u^\mu\big]
  =\varpi^{\rho\mu}\de_\mu\alpha_\rho
 =\f{1}{|\beta|}\underbrace{\varpi^{\rho\mu}}\ped{Antisym}\big( \underbrace{\varpi_{\rho\lambda}
 \varpi^\lambda_{\phantom{\mu}\mu}}\ped{Sym}+\underbrace{\alpha_\mu\alpha_\rho}\ped{Sym}\big)=0,
$$
where we have used the eq.~(\ref{gradalpha}). Another useful relation involving $\gamma$ is the contraction:
\begin{equation*}
 \gamma^\rho\varpi_{\rho\kappa}=\epsilon^{\rho\mu\nu\sigma}w_\mu\alpha_\nu u_\sigma\varpi_{\rho\kappa}
 =\epsilon^{\rho\mu\nu\sigma}\epsilon_{\rho\kappa\lambda\tau}w^\lambda u^\tau w_\mu \alpha_\nu u_\sigma 
 + \epsilon^{\rho\mu\nu\sigma}\alpha_\rho u_\kappa w_\mu \alpha_\nu u_\sigma-
 \epsilon^{\rho\mu\nu\sigma}\alpha_\kappa u_\rho w_\mu \alpha_\nu u_\sigma,
\end{equation*}
where we have used the decomposition (\ref{vortdec}). Then, because of the Levi-Civita tensor properties we find:
\begin{equation}\label{gammavarpi}
 \gamma^\rho\varpi_{\rho\kappa}=w^2\alpha_\kappa-(\alpha\cdot w)w_\kappa.
\end{equation}
We also need the derivative of the transverse projector
$\Delta$:
\begin{equation*}
 \de_\mu \Delta_{\rho\sigma}=\de_\mu\big( g_{\rho\sigma}-u_\rho u_\sigma\big)
 =-u_\sigma\de_\mu u_\rho - u_\rho \de_\mu u_\sigma
\end{equation*}
whence:
\begin{equation}\label{divdelta}
 \de^\rho \Delta_{\rho\sigma}=-u_\sigma \de^\rho u_\rho - u_\rho \de^\rho u_\sigma
 =-\f{\alpha_\sigma}{|\beta|}.
\end{equation}
Finally, we observe that
$$ 
  D \alpha^2 = D w^2 = 0
$$
as they are scalar functions.

To calculate the divergence of the stress-energy tensor we need to work out some intermediate relations
involving the derivatives of the four-vectors $u,\alpha,w,\gamma$. The first relation can be obtained by using the (\ref{divdelta}) and (\ref{dalpha2}):
\begin{equation}\label{relneeded1}
 \de^\mu\left(\alpha^2\Delta_{\mu\nu}\right)=\alpha^2\de^\mu\Delta_{\mu\nu}+\Delta_{\mu\nu}\de^\mu\alpha^2=
 - \f{\alpha^2}{|\beta|} \alpha_\nu + \partial_\nu \alpha^2 = 
 \frac{1}{|\beta|}\left((2w^2-\alpha^2)\alpha_\nu-2(\alpha\cdot w)w_\nu\right).
\end{equation}
The second by using the (\ref{divdelta}) and (\ref{dw2}):
\begin{equation}\label{relneeded2}
 \de^\mu\left(w^2\Delta_{\mu\nu} \right)=\Delta_{\mu\nu}\de^\mu w^2+w^2\de^\mu\Delta_{\mu\nu}
 =\frac{2}{|\beta|}\left(w^2\alpha_\nu-(\alpha\cdot w)w_\nu\right)-\frac{w^2\alpha_\nu}{|\beta|}
 =\frac{1}{|\beta|}\left(w^2\alpha_\nu-2(\alpha\cdot w)w_\nu\right).
\end{equation}
Moreover with~(\ref{gradalpha2}) and orthogonality properties:
\begin{equation}\label{relneeded3}
 \alpha_\mu\de^\mu\alpha_\nu  =\f{\alpha_\mu}{|\beta|}\left(-w^\mu w_\nu+\gamma^\mu u_\nu-u^\mu \gamma_\nu-(w^2+\alpha^2)u^\mu u_\nu+g^\mu_{\,\nu}w^2\right)
  =\f{w^2\alpha_\nu-(\alpha\cdot w)w_\nu}{|\beta|};
\end{equation}
and, using (\ref{dw}):
\begin{equation}\label{relneeded4}
 w_\mu \de^\mu w_\nu=\frac{w_\mu}{|\beta|}\left(-g^\mu_{\hphantom{\mu}\nu}(\alpha\cdot w)+\alpha^\mu    w_\nu\right)=
 \frac{1}{|\beta|}\left(-w_\nu (\alpha\cdot w)+(\alpha\cdot w)w_\nu\right)=0.
\end{equation}
Futhermore, taking advantage of (\ref{gradu}) and (\ref{gammavarpi}):
\begin{equation}\label{relneeded5}
 \gamma_\mu \de^\mu u_\nu=\f{\gamma_\mu}{|\beta|}(\varpi_{\nu}^{\,\mu}+\alpha^\mu u_\nu)=-\f{\gamma^\rho\varpi_{\rho\nu}}{|\beta|}
 =\f{(\alpha\cdot w)w_\nu-w^2\alpha_\nu}{|\beta|}.
\end{equation}
The last needed relation involves the gradients of $\gamma$ in eq.~(\ref{gamma}). By expanding its
definition and using the previous eqs.~(\ref{gradu},\ref{gradalpha2},\ref{dw}):
\begin{equation}\label{relneeded6}
\begin{split}
u_\mu\de^\mu\gamma_\nu &=\epsilon_{\nu\rho\sigma\tau}u_\mu\de^\mu\left(w^\rho\alpha^\sigma u^\tau\right)\\
&=\epsilon_{\nu\rho\sigma\tau}u_\mu(\de^\mu w^\rho)\alpha^\sigma u^\tau
+\epsilon_{\nu\rho\sigma\tau}u_\mu w^\rho(\de^\mu\alpha^\sigma) u^\tau
+\epsilon_{\nu\rho\sigma\tau}u_\mu w^\rho\alpha^\sigma \de^\mu u^\tau\\
&=-\epsilon_{\nu\rho\sigma\tau}u_\rho\alpha^\sigma u^\tau
-\epsilon_{\nu\rho\sigma\tau} w^\rho(\gamma^\sigma-\alpha^2 u^\sigma) u^\tau
+\epsilon_{\nu\rho\sigma\tau} w^\rho \alpha^\sigma \alpha^\tau\\
&=-\epsilon_{\nu\rho\sigma\tau} w^\rho\gamma^\sigma u^\tau=-\epsilon_{\nu\rho\sigma\tau}\epsilon^{\sigma\kappa\lambda\pi} w^\rho u^\tau w_\kappa u_\pi
=\frac{1}{|\beta|}(w_\nu (\alpha\cdot w) -\alpha_\nu w^2 -\alpha_\nu\alpha^2),
\end{split}
\end{equation}
where the known contractions of Levi-Civita tensor have been employed in the last
equality.

We are now in a position to enforce the continuity equation $\de^\mu T_{\mu\nu} = 0$, to the 
expression of the stress energy tensor~(\ref{setapp}). The first term involving energy density 
gives rise to:
\begin{equation}\label{eq:partiInizio}
\de^\mu\big(\rho \,u_\mu u_\nu\big) =u_\mu \big(\de^\mu \rho\big)u_\nu + \rho\, u_\nu\, \de^\mu u_\mu + \rho\, u_\mu\de^\mu u_\nu
   =\rho \f{\alpha_\nu}{|\beta|}
\end{equation}
using eqs.~(\ref{gradscalar}), (\ref{divu}) and (\ref{comderu}). Then, keeping in mind
the (\ref{df0}) which implies that $D U_\alpha = D U_w = \ldots = 0$ for all coefficients,
and using~(\ref{gradscalar}), (\ref{divu}), (\ref{comderu}), as well as $D\alpha^2=D w^2=0$:
\begin{gather}
  \de^\mu\big(U_\alpha\, \alpha^2 u_\mu u_\nu\big)=D U_\alpha \, \alpha^2 u_\nu 
  + U_\alpha\Big[\alpha^2 u_\nu\de^\mu u_\mu+u_\nu D \alpha^2+\alpha^2 u_\mu\de^\mu u_\nu\Big]=
  U_\alpha \f{\alpha^2\,\alpha_\nu}{|\beta|};\\
  \de^\mu\big(U_w\, w^2 u_\mu u_\nu\big)= D U_w \, w^2 u_\nu + 
  U_w\Big[w^2 u_\nu\de^\mu u_\mu+u_\nu Dw^2+w^2 u_\mu\de^\mu u_\nu\Big]=
  U_w \f{w^2\,\alpha_\nu}{|\beta|}.
\end{gather}
Now, using eqs.~(\ref{gradscalar}) and (\ref{divdelta}):
\begin{equation}
 \de^\mu\big( p\,\Delta_{\mu\nu}\big) =\Delta_{\mu\nu}\de^\mu p+p\de^\mu\Delta_{\mu\nu}
   =-\Big(p+|\beta|\f{\de p}{\de|\beta|}\Big)\f{\alpha_\nu}{|\beta|};
\end{equation}
and, thanks to eqs.~(\ref{gradscalar}) and~(\ref{relneeded1}):
\begin{equation}
 \de^\mu\big( D_\alpha\,\alpha^2\Delta_{\mu\nu}\big)=\Delta_{\mu\nu}\alpha^2\de^\mu 
 D_\alpha+D_\alpha\de^\mu\big(\alpha^2\Delta_{\mu\nu}\big)=
 -\Big(|\beta|\f{\de D_\alpha}{\de|\beta|}+D_\alpha \Big)\f{\alpha^2\alpha_\nu}{|\beta|}
 +2D_\alpha\f{w^2\alpha_\nu}{|\beta|}-2D_\alpha\f{(\alpha\cdot w)w_\nu}{|\beta|};
\end{equation}
likewise, because of (\ref{gradscalar}) and~(\ref{relneeded2}):
\begin{equation}
 \de^\mu\big(D_w w^2\Delta_{\mu\nu}\big)=w^2 \Delta_{\mu\nu} \de^\mu D_w + D_w \de^\mu \big( w^2 \Delta_{\mu\nu}\big)
 =\Big(D_w-|\beta|\f{\de D_w}{\de|\beta|}\Big)\f{w^2\alpha_\nu}{|\beta|}-2D_w\f{(\alpha\cdot w)w_\nu}{|\beta|}.
\end{equation}
Furthermore, using (\ref{gradscalar}), (\ref{divu}) and (\ref{relneeded3}):
\begin{equation}
 \de^\mu \big( A\, \alpha_\mu \alpha_\nu\big)=
  \alpha_\mu \alpha_\nu \big( \de^\mu A\big)+A\,\alpha_\nu\de^\mu\alpha_\mu+A\,\alpha_\mu\de^\mu\alpha_\nu=
  -\big( |\beta|\f{\de A}{\de|\beta|}+A\big)\f{\alpha^2\alpha_\nu}{|\beta|}
  +3A\f{w^2\alpha_\nu}{|\beta|}-A\f{(\alpha\cdot w)\alpha_\nu}{|\beta|}
\end{equation}
and, similarly, by means of (\ref{gradscalar}), (\ref{divw}) and (\ref{relneeded4}):
\begin{equation}
 \de^\mu\big(W w_\mu w_\nu\big)=(\de^\mu W)w_\mu w_\nu+Ww_\nu\de^\mu w_\mu+Ww_\mu\de^\mu w_\nu
 =-\Big(|\beta|\f{\de W}{\de|\beta|}+3W\Big)\f{(\alpha\cdot w) w_\nu}{|\beta|}.
\end{equation}
Finally, using again (\ref{gradscalar}) and the relations (\ref{relneeded5}) and (\ref{relneeded6}) 
we can determine the term involving the gradients of $\gamma$:
\begin{equation}\label{eq:partiFine}
 \de^\mu\left[ G (u_\mu \gamma_\nu+u_\nu \gamma_\mu)\right]=(u_\mu \gamma_\nu+u_\nu \gamma_\mu)\de^\mu G+
 G u_\mu \de^\mu \gamma_\nu + G\gamma_\mu \de^\mu u_\nu
=2G\f{(\alpha\cdot w)w_\nu}{|\beta|}-2G\f{w^2\alpha_\nu}{|\beta|}.
\end{equation}

The divergence of the stress-energy tensor is obtained by summing all right-hand sides of the 
eqs.~(\ref{eq:partiInizio}-\ref{eq:partiFine}). As it can be readily checked, the resulting 
expression involves the sum of terms multiplying $\alpha_\nu,\,\alpha^2\alpha_\nu,\,w^2\alpha_\nu$ 
and $(\alpha\cdot w)w_\nu$. As the divergence should vanish independently of the vectors 
$w$ and $\alpha$, which have as many degrees of freedom as the thermal vorticity (that is 6),
the conclusion is that each coefficient of the above combinations must be zero. As a result, 
four equations are obtained. The coefficient of $\alpha_\nu$ gives rise to:
\begin{equation*}
 \rho+p+|\beta|\frac{\de p}{\de|\beta|}\Big|_\zeta=0,
\end{equation*}
which is but the well known thermodynamic relation between energy density and pressure. 
The other three coefficients yield three conditions on the second-order coefficients:
\begin{equation*}
 U_\alpha=-|\beta|\f{\de}{\de|\beta|}\big(D_\alpha+A\big)-\big(D_\alpha+A\big);
\end{equation*}
\begin{equation}
 \label{eq:partialresult1}
 U_w=-|\beta|\f{\de}{\de|\beta|}D_w+(2D_\alpha+D_w+3A)-2G;
\end{equation}
and:
\begin{equation}
 \label{eq:partialresult2}
 2G=2\big(D_\alpha+D_w\big)+A+|\beta|\f{\de}{\de|\beta|}W+3W.
\end{equation}
Subtracting eq.~(\ref{eq:partialresult2}) from eq.~(\ref{eq:partialresult1}) we can 
simplify some terms and obtain:
\begin{equation*}
 U_w=-|\beta|\f{\de}{\de|\beta|}\big(D_w+W\big)-D_w+2A-3W,
\end{equation*}
which completes the proof of the relations in eq.~(\ref{coeffrel}).
%



\end{document}